\documentclass{aa}  
\usepackage{graphicx}
\usepackage{txfonts}
\newcommand{\mincir}{\raise -2.truept\hbox{\rlap{\hbox{$\sim$}}\raise5.truept
\hbox{$<$}\ }}
\newcommand{\magcir}{\raise -2.truept\hbox{\rlap{\hbox{$\sim$}}\raise5.truept
\hbox{$>$}\ }}
\newcommand{\siml}{\raise -2.truept\hbox{\rlap{\hbox{$\sim$}}\raise5.truept
\hbox{$<$}\ }}
\newcommand{\simg}{\raise -2.truept\hbox{\rlap{\hbox{$\sim$}}\raise5.truept
\hbox{$>$}\ }}
\newcommand{\be}{\begin{equation}}
\newcommand{\ee}{\end{equation}}
\newcommand{\ba}{\begin{eqnarray}}
\newcommand{\ea}{\end{eqnarray}}
\newcommand {\kpc} {$h_{70}^{-1}$ kpc $\;$}
\newcommand {\kpcc} {$h_{70}^{-1}$ kpc}

\newcommand {\h} {$h_{70}^{-1}$ Mpc$\;$}
\newcommand {\hh} {$h_{70}^{-1}$ Mpc}
\newcommand {\hhh} {\;h_{70}^{-1} \mathrm{Mpc}}
\newcommand {\ks} {km~s$^{-1} \;$}
\newcommand {\kss} {km~s$^{-1}$}

\newcommand {\mqui} {$\times 10^{15}\;h_{70}^{-1}\;M_{\odot} \;$}
\newcommand {\mquii} {$\times 10^{15}\;h_{70}^{-1}\;M_{\odot}$}

\newcommand{\degree}{\ensuremath{\mathrm{^\circ}}}
\newcommand{\arcm}{\ensuremath{\mathrm{^\prime}\;}}
\newcommand{\arcs}{\ensuremath{\arcmm\hskip -0.1em\arcmm \;}}
\newcommand{\arcmm}{\ensuremath{\mathrm{^\prime}}}
\newcommand{\arcss}{\ensuremath{\arcmm\hskip -0.1em\arcmm}}

\newcommand{\dotsec}{\,\rlap{\hbox{$\mathrm{^s}$}}{\hbox{$.$}}\,}

\begin{document}
   \title{Internal dynamics of Abell 2294: a massive, likely
     merging cluster}

   \author{
M. Girardi\inst{1,2}
          \and
W. Boschin\inst{3}
          \and
R. Barrena\inst{4,5}
}

   \offprints{M. Girardi, \email{girardi@oats.inaf.it}}

   \institute{ 
Dipartimento di Fisica dell' Universit\`a degli Studi
     di Trieste - Sezione di Astronomia, via Tiepolo 11, I-34143
     Trieste, Italy\\ 
\and INAF - Osservatorio Astronomico di Trieste,
     via Tiepolo 11, I-34143 Trieste, Italy\\ 
\and Fundaci\'on Galileo
     Galilei - INAF, Rambla Jos\'e Ana Fern\'andez Perez 7, E-38712
     Bre\~na Baja (La Palma), Canary Islands, Spain\\ 
\and Instituto de
     Astrof\'{\i}sica de Canarias, C/V\'{\i}a L\'actea s/n, E-38205 La
     Laguna (Tenerife), Canary Islands, Spain\\ 
\and Departamento de
     Astrof\'{\i}sica, Universidad de La Laguna, Av. del
     Astrof\'{\i}sico Franciso S\'anchez s/n, E-38205 La Laguna
     (Tenerife), Canary Islands, Spain
}

\date{Received  / Accepted }

\abstract{The mechanisms giving rise to diffuse radio emission in
  galaxy clusters, and in particular their connection with cluster
  mergers, are still debated.}{We aim to obtain new insights into the
  internal dynamics of the cluster Abell 2294, recently shown to host
  a radio halo.}  {Our analysis is mainly based on redshift data for
  88 galaxies acquired at the Telescopio Nazionale Galileo. We combine
  galaxy velocities and positions to select 78 cluster galaxies and
  analyze its internal dynamics. We also use new photometric data
  acquired at the Isaac Newton Telescope and X--ray data from the
  Chandra archive.}{We re--estimate the redshift of the large,
  brightest cluster galaxy (BCG) obtaining $\left<z\right>=0.1690$,
  well at rest within the cluster. We estimate a quite large
  line--of--sight (LOS) velocity dispersion $\sigma_{\rm V}\sim 1400$
  \ks and X--ray temperature $T_{\rm X}\sim \,$10 keV. Our
    optical and X--ray analysis detects evidence for substructure.
    Our results are consistent with the presence of two massive
    subclusters separated by a LOS rest frame velocity difference
    $V_{\rm rf}\sim 2000$ \kss, very closely projected in the plane of
    sky along the SE--NW direction.  The observational picture,
    interpreted through the analytical two--body model, suggests that
    A2294 is a cluster merger elongated mainly in the LOS direction
    and catched during the bound outgoing phase, a few fractions of
    Gyr after the core crossing.  We find Abell 2294 is a very massive
    cluster with a range of $M=2-4$ \mquii, depending on the adopted
    model.  Moreover, contradicting previous findings, our new data
  do exclude the presence of the H$\alpha$ emission in the spectrum of
  the BCG galaxy.}{The outcoming picture of Abell 2294 is that of a
  massive, quite ``normal'' merging cluster, as found for many
  clusters showing diffuse radio sources. However, maybe due to the
  particular geometry, more data are needed for a definitive,
    more quantitative conclusion.}


   \maketitle
%

\section{Introduction}
\label{intr}

Merging processes constitute an essential ingredient of the evolution
of galaxy clusters (see Feretti et al. \cite{fer02b} for a review). An
interesting aspect of these phenomena is the possible connection of
cluster mergers with the presence of extended, diffuse radio sources:
halos and relics. The synchrotron radio emission of these sources
demonstrates the existence of large--scale cluster magnetic fields and
of widespread relativistic particles. Cluster mergers have been
suggested to provide the large amount of energy necessary for electron
reacceleration up to relativistic energies and for magnetic field
amplification (Tribble \cite{tri93}; Feretti \cite{fer99}; Feretti
\cite{fer02a}; Sarazin \cite{sar02}). Radio relics (``radio gischts''
as referred by Kempner et al. \cite{kem04}), which are polarized and
elongated radio sources located in the cluster peripheral regions,
seem to be directly associated with merger shocks (e.g., Ensslin et
al. \cite{ens98}; Roettiger et al. \cite{roe99}; Ensslin \&
Gopal--Krishna \cite{ens01}; Hoeft et al. \cite{hoe04}).  Radio halos,
unpolarized sources which permeate the cluster volume similarly to the
X--ray emitting gas (intracluster medium, hereafter ICM), are more
likely to be associated with the turbulence following a cluster merger
(Cassano \& Brunetti \cite{cas05}; Brunetti et
al. \cite{bru09}). However, the precise radio halos/relics formation
scenario is still debated since the diffuse radio sources are quite
uncommon and only recently one can study these phenomena on the basis
of a sufficient statistics (few dozen clusters up to $z\sim 0.3$,
e.g., Giovannini et al. \cite{gio99}; see also Giovannini \& Feretti
\cite{gio02}; Feretti \cite{fer05}; Giovannini et al. \cite{gio09})
and attempt a classification (e.g., Kempner et al. \cite{kem04};
Ferrari et al. \cite{ferr08}). It is expected that new telescopes will
largely increase the statistics of diffuse sources (e.g. LOFAR,
Cassano et al. \cite{cas09}).

From the observational point of view, there is growing evidence of the
connection between diffuse radio emission and cluster merging, since
up to now diffuse radio sources have been detected only in merging
systems. In several cases the cluster dynamical state has been derived
from X--ray observations (see Buote \cite{buo02}; Feretti \cite{fer06}
and \cite{fer08} and refs. therein).  Optical data are a powerful way
to investigate the presence and the dynamics of cluster mergers (e.g.,
Girardi \& Biviano \cite{gir02}), too. The spatial and kinematical
analysis of member galaxies allow us to detect and measure the amount
of substructure, to identify and analyze possible pre--merging clumps
or merger remnants.  This optical information is really complementary
to X--ray information since galaxies and intra--cluster medium react
on different time scales during a merger (see, e.g., numerical
simulations by Roettiger et al. \cite{roe97}).

In this context we are conducting an intensive observational and data
analysis program to study the internal dynamics of clusters with
diffuse radio emission by using member galaxies (Girardi et
al. \cite{gir07}\footnote{see also the web site of the DARC (Dynamical
  Analysis of Radio Clusters) project:
  http://adlibitum.oat.ts.astro.it/girardi/darc.}). Most clusters
showing diffuse radio emission have a large gravitational mass (larger
than $0.7\times 10^{15}$ within $2$ \hh; see Giovannini \& Feretti
\cite{gio02}) and, indeed, most clusters we analyzed are very massive
clusters with few exceptions (Boschin et al. \cite{bos08}).

During our observational program we have conducted an intensive study
of the cluster \object{Abell 2294} (hereafter A2294).

\begin{figure*}
\centering 
\includegraphics[width=18cm]{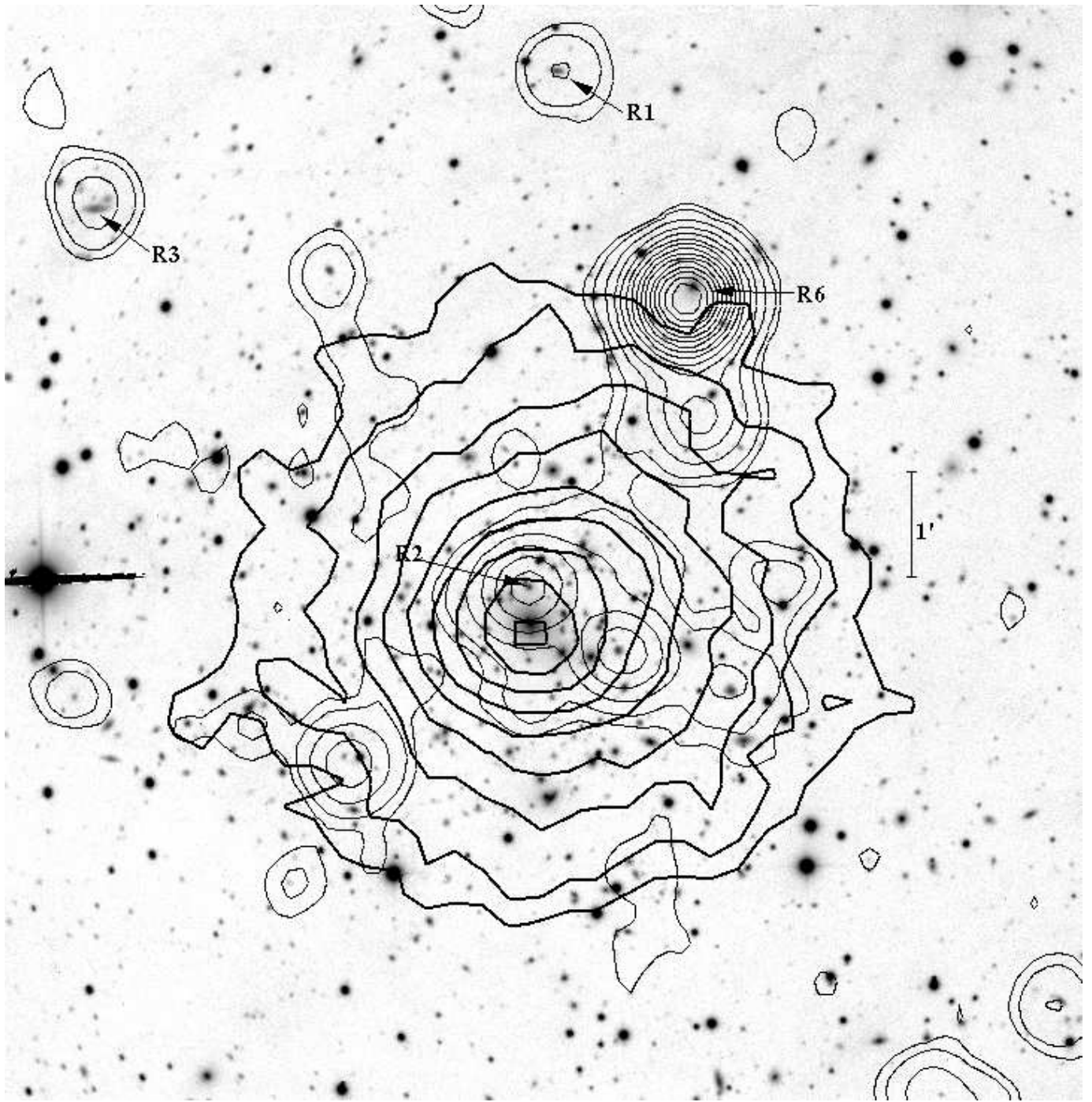}
\caption{INT $R$--band image of the cluster A2294 (North at the top
  and East to the left) with, superimposed, the contour levels of the
  Chandra archival image ID~3246 (thick contours; photons in the
  energy range 0.5--2 keV) and the contour levels of a VLA radio image
  at 1.4 GHz (thin contours, see Giovannini et
  al. \cite{gio09}). Labels and arrows highlight the positions of
  radio sources listed by Rizza et al. (\cite{riz03}).}
\label{figimage1}
\end{figure*}

A2294 is a very rich, X--ray luminous, and hot Abell cluster: Abell
richness class $=2$ (Abell et al. \cite{abe89});
$L_\mathrm{X}$(0.1--2.4 keV)=6.6$\times 10^{44} \ h_{50}^{-2}$
erg\ s$^{-1}$; $T_{\rm X}=\,$8-9 keV recovered from ROSAT and Chandra data
(Ebeling et al. \cite{ebe98}; Rizza et al. \cite{riz98}; 
Maughan et al. \cite{mau08}). Optically,
the cluster is classified as Bautz--Morgan class II (Abell et
al. \cite{abe89}) and is dominated by a central, large brightest
cluster galaxy (BCG, see Fig.~\ref{figimage1}).

From both ROSAT and Chandra data A2294 is known for having no cool
core (Rizza et al. \cite{riz98}; Bauer et al. \cite{bau05}). As for
the presence of possible substructure, using ROSAT data, Rizza et
al. (\cite{riz98}) found evidence of a centroid shift and detect a
Southern excess in the X--ray emission. Moreover, using Chandra data,
Hashimoto et al. (\cite{has07}) classified A2294 as a ``distorted''
cluster due to its large value of the asymmetry parameter. Indeed,
A2294 is a very peculiar cluster since, in contrast with the absence
of a cooling core, it is very compact in its X-ray appearance (see
Fig.3 of Bauer et al. \cite{bau05}). Out of a sample of 115
  clusters recently analyzed using Chandra data, A2294 is one with the
  smallest ellipticity, while shows a not large, but highly
  significant centroid shift (Maughan et al. \cite{mau08}).

A2294 is also peculiar for another aspect.  Out of a sample of 13
clusters at $z\sim 0.15-0.4$ showing evidence for H$\alpha$ emission
in the BCG spectrum, it is the only one not showing a cool core (see
Fig.5 of Bauer et al. \cite{bau05}). The correlation between BCG
H$\alpha$ emission and the presence of a cool core is also true for
nearby clusters where `` H$\alpha$ luminous galaxies lie at the center
of large cool cores, although this special cluster environment does
not guarantee the emission-line nebulosity in its BCG'' (Peres et
al. \cite{per98}).  More recent observations also agree that H$\alpha$
emission is more typical of cool core clusters than of non--cool core
clusters ($\sim 70\%$ against $\sim 10\% $, Edwards et
al. \cite{edw07}).

As for the diffuse radio emission, Owen et al. (\cite{owe99}) first
reported the existence of a detectable diffuse radio source in this
cluster. Despite of the presence of some disturbing pointlike sources
in the central region of the cluster, Giovannini et al. (\cite{gio09})
could detect a radio--halo 3\arcm in size. In particular, the position
of A2294 in the $P_{1.4\,\rm{GHz}}$ (radio power at 1.4 GHz) - $L_{\rm
  X}$ plane is consistent with that of all other radio--halo clusters
(see Fig.~17 of Giovannini et al. \cite{gio09}).

To date poor optical data are available. The cluster redshift reported
in the literature ($z=0.178$) is only based on the BCG H$\alpha$
emission line (Crawford et al. \cite{cra95}). Instead, the real
cluster redshift, as estimated in this paper, is rather
$\left<z\right>=0.169$ fully consistent with that measured on the BCG
on the base of our data which, indeed, do not show any evidence of
H$\alpha$ emission (see \S~\ref{data}).

Our new spectroscopic and photometric data come from the Telescopio
Nazionale Galileo (TNG) and the Isaac Newton Telescope (INT),
respectively. Our present analysis is based on these optical data and
X--ray Chandra archival data.

This paper is organized as follows. We present our new optical data
and the cluster catalog in Sect.~2. We present our results about the
cluster structure based on optical and X--ray data in Sects.~3 and 4,
respectively.  We briefly discuss our results and give our conclusions
in Sect.~5.

Unless otherwise stated, we give errors at the 68\% confidence level
(hereafter c.l.).

Throughout this paper, we use $H_0=70$ km s$^{-1}$ Mpc$^{-1}$ in a
flat cosmology with $\Omega_0=0.3$ and $\Omega_{\Lambda}=0.7$. In the
adopted cosmology, 1\arcm corresponds to $\sim 173$ \kpc at the
cluster redshift.

\begin{figure*}
\centering 
\includegraphics[width=18cm]{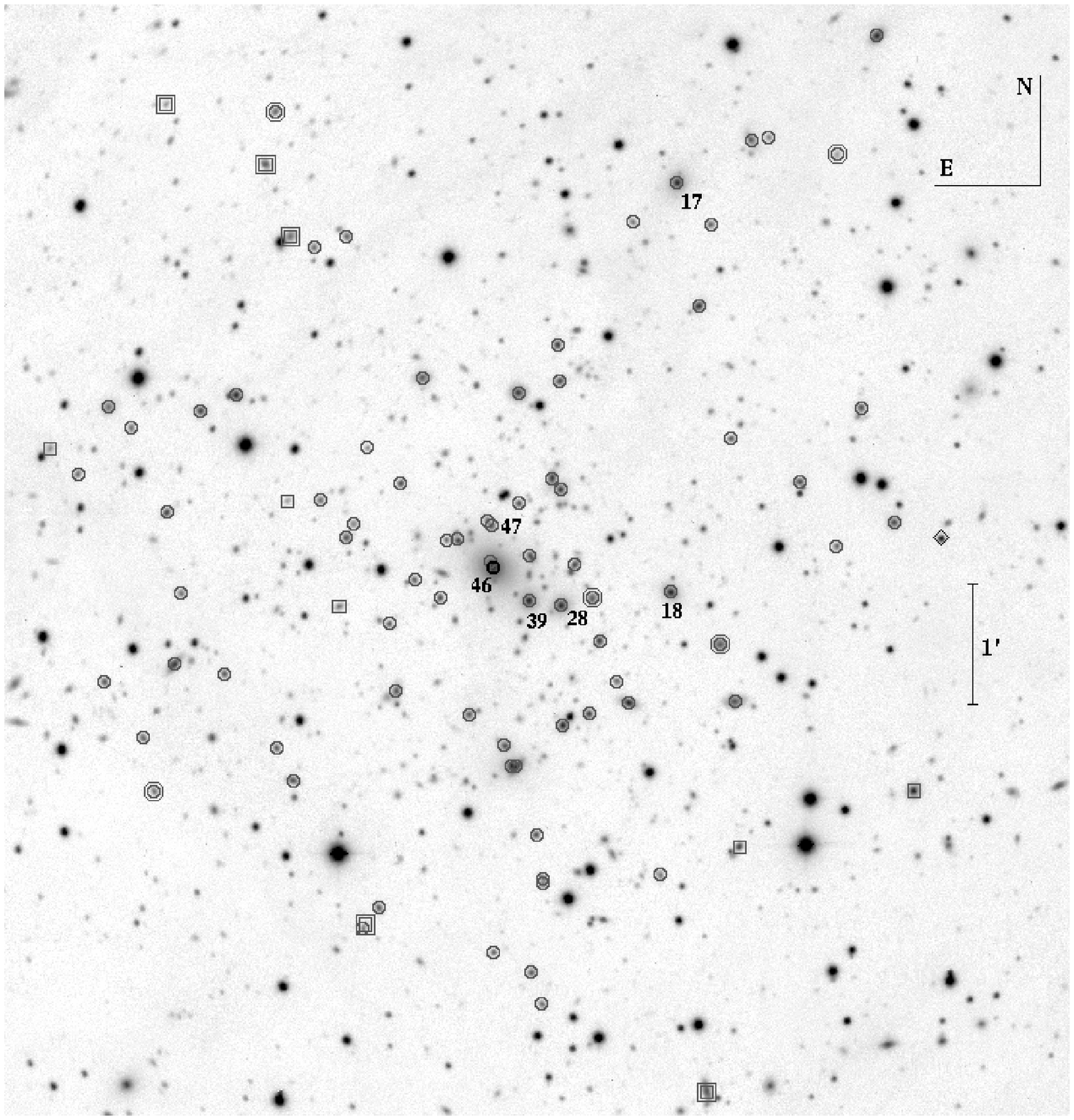}
\caption{INT $R$--band image of the cluster A2294 (North at the top
  and East to the left). Circles and squares indicate cluster members
  and non--members, respectively (see Table~\ref{catalogA2294}). Solid
  circle in the center highlights the position of the BCG galaxy. 
    Annuli and box annuli show member and non--member emission line
    galaxies, respectively. Labels indicate the IDs of cluster
  galaxies cited in the text. A diamond at the right border of the
  image highlights a QSO at $z\sim$2.1.} 
\label{figottico2}
\end{figure*}

\section{New data and galaxy catalog}
\label{data}

Multi--object spectroscopic observations of A2294 were carried out at
the TNG telescope in December 2007 and August 2008. We used
DOLORES/MOS with the LR--B Grism 1, yielding a dispersion of 187
\AA/mm. We used the new $2048\times2048$ pixels E2V CCD, with a pixel
size of 13.5 $\mu$m. In total we observed 4 MOS masks (3 in 2007 and 1
in 2008) for a total of 124 slits. We acquired three exposures of 1800
s for each mask. Wavelength calibration was performed using
Helium--Argon lamps. Reduction of spectroscopic data was carried out
with the IRAF\footnote{IRAF is distributed by the National Optical
  Astronomy Observatories, which are operated by the Association of
  Universities for Research in Astronomy, Inc., under cooperative
  agreement with the National Science Foundation.} package. Radial
velocities were determined using the cross--correlation technique
(Tonry \& Davis \cite{ton79}) implemented in the RVSAO package
(developed at the Smithsonian Astrophysical Observatory Telescope Data
Center). Each spectrum was correlated against six templates for a
variety of galaxy spectral types: E, S0, Sa, Sb, Sc, Ir (Kennicutt
\cite{ken92}). The template producing the highest value of $\cal R$,
i.e., the parameter given by RVSAO and related to the
signal--to--noise ratio of the correlation peak, was chosen. Moreover,
all spectra and their best correlation functions were examined
visually to verify the redshift determination. In six cases (IDs.~5,
13, 15, 60, 81 and 82; see Table~\ref{catalogA2294}) we took the EMSAO
redshift as a reliable estimate of the redshift.

Our spectroscopic survey in the field of A2294 consists of spectra for
88 galaxies.


\begin{table}[!ht]
        \caption[]{Velocity catalog of 88 spectroscopically measured
          galaxies in the field of the cluster A2294. ID~46, in
          boldface highlights the BCG. IDs in italics are
          non-member galaxies.}
         \label{catalogA2294}
              $$ 
           \begin{array}{r c c c r r l}
            \hline
            \noalign{\smallskip}
            \hline
            \noalign{\smallskip}

\mathrm{ID} & \mathrm{\alpha},\mathrm{\delta}\,(\mathrm{J}2000)  & B & R &\mathrm{v}\,\,\,\,\,\,\, & \mathrm{\Delta}\mathrm{v}& \mathrm{Emission}\\
  & & & &\mathrm{(\,km}&\mathrm{s^{-1}\,)}& \mathrm{lines}\\
            \hline
            \noalign{\smallskip}

\textit{1}   &17\ 21\ 01.75 ,+85\ 51\ 21.1&18.65&    17.21& 28640&  57&                    \\
 2           &17\ 21\ 09.36 ,+85\ 53\ 32.7&19.79&    17.82& 51333&  67&                    \\
 3           &17\ 21\ 14.69 ,+85\ 57\ 31.3&19.04&    16.92& 52128&  55&                    \\
 4           &17\ 21\ 23.83 ,+85\ 54\ 29.0&20.24&    18.17& 51318&  77&                    \\
 5           &17\ 21\ 33.77 ,+85\ 56\ 33.5&20.29&    19.42& 48560&  16& \mathrm{[OII]}     \\
 6           &17\ 21\ 35.57 ,+85\ 53\ 21.4&20.89&    18.88& 50550&  71&                    \\
 7           &17\ 21\ 52.15 ,+85\ 53\ 53.1&19.92&    18.05& 52312&  59&                    \\
 8           &17\ 22\ 05.21 ,+85\ 56\ 42.0&20.75&    19.10& 50779& 147&                    \\
 9           &17\ 22\ 13.44 ,+85\ 56\ 40.9&19.86&    17.83& 52063&  59&                    \\
\textit{10}  &17\ 22\ 20.83 ,+85\ 50\ 54.7&19.42&    18.19& 36167&  77&                    \\
11           &17\ 22\ 22.49 ,+85\ 52\ 05.8&18.51&    17.21& 50279& 109&                    \\
12           &17\ 22\ 23.50 ,+85\ 54\ 14.6&20.77&    18.81& 50787&  66&                    \\
13           &17\ 22\ 29.06 ,+85\ 52\ 33.8&18.49&    17.68& 53150&  75& \mathrm{[OII],H\beta} \\
14           &17\ 22\ 32.31 ,+85\ 55\ 59.5&20.97&    18.83& 51408& 112&                    \\
\textit{15}  &17\ 22\ 35.95 ,+85\ 48\ 54.6&18.56&    17.22& 21565& 109& \mathrm{H\alpha}   \\
16           &17\ 22\ 37.99 ,+85\ 55\ 19.7&19.55&    17.54& 49334&  52&                    \\
17           &17\ 22\ 47.69 ,+85\ 56\ 20.1&18.87&    16.64& 49774&  62&                    \\
18           &17\ 22\ 51.26 ,+85\ 52\ 59.6&18.99&    16.89& 51816&  45&                    \\
19           &17\ 22\ 56.69 ,+85\ 50\ 41.1&20.57&    18.79& 49831&  77&                    \\
20           &17\ 23\ 08.14 ,+85\ 56\ 01.1&21.08&    19.14& 50558&  77&                    \\
21           &17\ 23\ 10.54 ,+85\ 52\ 05.5&19.02&    17.04& 50342&  45&                    \\
22           &17\ 23\ 15.98 ,+85\ 52\ 16.1&20.84&    18.94& 48439& 108&                    \\
23           &17\ 23\ 23.83 ,+85\ 52\ 35.9&19.98&    17.89& 54000&  55&                    \\
24           &17\ 23\ 26.93 ,+85\ 52\ 56.8&19.06&    18.06& 53404& 109& \mathrm{[OII],H\beta}, \\
             &                            &     &         &      &    & \mathrm{H\alpha,[NII]} \\
25           &17\ 23\ 28.63 ,+85\ 52\ 00.2&19.92&    17.97& 52231&  60&                    \\
26           &17\ 23\ 35.18 ,+85\ 53\ 13.3&20.05&    17.97& 50231&  54&                    \\
27           &17\ 23\ 40.37 ,+85\ 51\ 54.6&19.16&    17.36& 50708&  66&                    \\
28           &17\ 23\ 41.37 ,+85\ 52\ 53.2&18.76&    16.64& 49068&  34&                    \\
29           &17\ 23\ 41.47 ,+85\ 53\ 50.2&19.54&    17.60& 48114&  64&                    \\
30           &17\ 23\ 41.73 ,+85\ 54\ 43.2&19.97&    17.91& 52330&  48&                    \\
31           &17\ 23\ 42.82 ,+85\ 55\ 01.1&20.22&    18.09& 47906&  70&                    \\
32           &17\ 23\ 45.34 ,+85\ 53\ 55.6&19.33&    17.39& 49027&  52&                    \\
33           &17\ 23\ 49.32 ,+85\ 50\ 37.0&20.67&    18.87& 51693& 113&                    \\
34           &17\ 23\ 49.37 ,+85\ 50\ 39.7&20.12&    18.50& 53475&  57&                    \\
35           &17\ 23\ 49.78 ,+85\ 49\ 38.1&20.67&    18.70& 52794&  99&                    \\
36           &17\ 23\ 52.46 ,+85\ 51\ 01.1&20.48&    18.41& 53149&  59&                    \\
37           &17\ 23\ 54.51 ,+85\ 49\ 53.6&20.17&    18.34& 50294&  77&                    \\
38           &17\ 23\ 55.53 ,+85\ 53\ 17.8&19.82&    17.92& 51367& 104&                    \\
39           &17\ 23\ 55.92 ,+85\ 52\ 55.7&18.67&    16.76& 49221&  41&                    \\
40           &17\ 24\ 00.51 ,+85\ 53\ 43.4&20.52&    18.44& 52927&  78&                    \\
41           &17\ 24\ 00.91 ,+85\ 54\ 37.6&19.09&    17.10& 46553&  60&                    \\
42           &17\ 24\ 01.54 ,+85\ 51\ 34.3&18.16&    17.18& 52055&  31&                    \\
43           &17\ 24\ 03.53 ,+85\ 51\ 34.7&19.26&    17.80& 49876&  69&                    \\
44           &17\ 24\ 06.91 ,+85\ 51\ 44.7&20.41&    18.19& 51991&  69&                    \\

                        \noalign{\smallskip}			    
            \hline					    
            \noalign{\smallskip}			    
            \hline					    
         \end{array}
     $$ 
         \end{table}
\addtocounter{table}{-1}
\begin{table}[!ht]
          \caption[ ]{Continued.}
     $$ 
           \begin{array}{r c c c r r l}
            \hline
            \noalign{\smallskip}
            \hline
            \noalign{\smallskip}

\mathrm{ID} & \mathrm{\alpha},\mathrm{\delta}\,(\mathrm{J}2000)  & B & R &\mathrm{v}\,\,\,\,\,\,\, & \mathrm{\Delta}\mathrm{v}& \mathrm{Emission}\\
  & & & &\mathrm{(\,km}&\mathrm{s^{-1}\,)}& \mathrm{lines}\\

            \hline
            \noalign{\smallskip}

45           &17\ 24\ 11.76 ,+85\ 50\ 03.4&20.91&    19.00& 51059& 133&                    \\
\textbf{46}  &17\ 24\ 12.14 ,+85\ 53\ 12.0&17.12&    15.01& 50671&  66&                    \\
47           &17\ 24\ 12.94 ,+85\ 53\ 32.4&20.02&    18.05& 48983&  71&                    \\
48           &17\ 24\ 13.62 ,+85\ 53\ 14.7&20.03&    17.86& 51068& 186&                    \\
49           &17\ 24\ 14.94 ,+85\ 53\ 35.1&20.34&    18.58& 50477& 101&                    \\
50           &17\ 24\ 22.95 ,+85\ 52\ 00.0&20.06&    18.12& 52427&  97&                    \\
51           &17\ 24\ 28.34 ,+85\ 53\ 25.9&19.69&    17.61& 49638&  63&                    \\
52           &17\ 24\ 33.43 ,+85\ 53\ 25.5&21.04&    18.95& 50207&  74&                    \\
53           &17\ 24\ 36.29 ,+85\ 52\ 57.0&20.72&    18.58& 49332&  88&                    \\
54           &17\ 24\ 44.81 ,+85\ 54\ 44.7&19.57&    17.57& 51597&  76&                    \\
55           &17\ 24\ 48.05 ,+85\ 53\ 05.9&20.52&    18.68& 48339&  76&                    \\
56           &17\ 24\ 54.51 ,+85\ 53\ 53.1&20.52&    18.45& 49696&  70&                    \\
57           &17\ 24\ 55.99 ,+85\ 52\ 11.3&19.92&    17.84& 49614&  85&                    \\
58           &17\ 24\ 59.28 ,+85\ 52\ 44.2&20.67&    18.67& 49486&  56&                    \\
59           &17\ 25\ 03.41 ,+85\ 50\ 25.1&20.52&    18.42& 52052&  74&                    \\
\textit{60}  &17\ 25\ 09.21 ,+85\ 50\ 16.9&21.11&    20.61& 67190& 313& \mathrm{[OII],H\beta},\\
             &                            &     &         &      &    & \mathrm{[OIII]}    \\
61           &17\ 25\ 09.98 ,+85\ 54\ 10.7&21.77&    19.77& 49754& 104&                    \\
62           &17\ 25\ 10.31 ,+85\ 50\ 15.0&20.77&    18.77& 51675& 218&                    \\
63           &17\ 25\ 15.91 ,+85\ 53\ 33.0&19.87&    19.02& 53186&  66&                    \\
64           &17\ 25\ 19.18 ,+85\ 53\ 26.5&20.20&    18.21& 49892& 116&                    \\
65           &17\ 25\ 19.99 ,+85\ 55\ 54.0&20.51&    18.71& 51267& 113&                    \\
\textit{66}  &17\ 25\ 22.27 ,+85\ 52\ 52.7&20.31&    18.92&146216&  80&                    \\
67           &17\ 25\ 31.13 ,+85\ 53\ 44.8&20.58&    18.69& 51558&  57&                    \\
68           &17\ 25\ 34.97 ,+85\ 55\ 48.5&20.78&    18.84& 51215&  67&                    \\
69           &17\ 25\ 42.63 ,+85\ 51\ 27.3&20.08&    18.04& 52589&  70&                    \\
\textit{70}  &17\ 25\ 45.98 ,+85\ 55\ 53.9&19.95&    18.29& 58151&  92& \mathrm{[OII]}     \\
\textit{71}  &17\ 25\ 46.11 ,+85\ 53\ 44.3&21.58&    19.96& 81954& 294&                    \\
72           &17\ 25\ 50.09 ,+85\ 51\ 43.2&20.54&    18.58& 52040&  49&                    \\
73           &17\ 25\ 53.45 ,+85\ 56\ 55.0&19.77&    18.66& 48662&  85& \mathrm{[OII],H\beta}\\
\textit{74}  &17\ 25\ 57.79 ,+85\ 56\ 29.1&19.54&    18.27& 57854& 111& \mathrm{[OII],H\beta}\\
75           &17\ 26\ 10.20 ,+85\ 54\ 35.9&19.31&    17.25& 50852&  48&                    \\
76           &17\ 26\ 14.14 ,+85\ 52\ 19.3&20.50&    18.59& 50445&  66&                    \\
77           &17\ 26\ 26.37 ,+85\ 54\ 27.9&19.78&    17.83& 51769&  80&                    \\
78           &17\ 26\ 34.30 ,+85\ 52\ 58.7&20.37&    18.76& 47581&  87&                    \\
79           &17\ 26\ 36.99 ,+85\ 52\ 23.7&18.83&    17.04& 49923&  60&                    \\
80           &17\ 26\ 41.11 ,+85\ 53\ 38.2&19.93&    18.02& 49484&  64&                    \\
\textit{81}  &17\ 26\ 44.04 ,+85\ 56\ 58.2&21.52&    19.85& 89709&  18& \mathrm{[OII]}     \\
82           &17\ 26\ 45.34 ,+85\ 51\ 21.5&19.95&    18.93& 49890&  64& \mathrm{[OII],H\beta}, \\
             &                            &     &         &      &    & \mathrm{[OIII],H\alpha}\\ 
83           &17\ 26\ 50.47 ,+85\ 51\ 47.5&20.52&    18.71& 49422&  71&                    \\
84           &17\ 26\ 58.03 ,+85\ 54\ 19.1&20.69&    18.93& 52827&  97&                    \\
85           &17\ 27\ 08.45 ,+85\ 52\ 14.8&20.66&    18.60& 51085& 104&                    \\
86           &17\ 27\ 08.50 ,+85\ 54\ 29.4&19.54&    17.99& 48738&  83&                    \\
87           &17\ 27\ 22.08 ,+85\ 53\ 56.5&20.38&    18.82& 52246&  97&                    \\
\textit{88}  &17\ 27\ 35.04 ,+85\ 54\ 08.7&21.18&    18.98& 90810&  92&                    \\

                        \noalign{\smallskip}			    
            \hline					    
            \noalign{\smallskip}			    
            \hline					    
         \end{array}
     $$ 
\end{table}


The nominal errors as given by the cross--correlation are known to be
smaller than the true errors (e.g., Malumuth et al. \cite{mal92};
Bardelli et al. \cite{bar94}; Ellingson \& Yee \cite{ell94}; Quintana
et al. \cite{qui00}). Duplicate observations for the same
galaxy allowed us to estimate real intrinsic errors in data of the
same quality taken with the same instrument (e.g. Barrena et
al. \cite{bar07a}, \cite{bar07b}).  Here we have a limited number of
double determinations (i.e. five galaxies as coming from four
different masks) thus we decided to apply the correction already
applied in above studies. Hereafter we assume that true errors are
larger than nominal cross--correlation errors by a factor 1.4. For the
five galaxies with two redshift estimates we used the weighted mean of
the two measurements and the corresponding errors.

The median error on $cz$ is 71 \kss.  

As far as photometry is concerned, our observations were carried out
with the Wide Field Camera (WFC), mounted at the prime focus of the
2.5m INT telescope. We observed A2294 in May 18th 2007 with filters
$B_{\rm H}$ and $R_{\rm H}$ in photometric conditions and a seeing of
$\sim$1.5\arcss.

The WFC consists of a four--CCD mosaic covering a
33\arcmm$\times$33\arcm field of view, with only a 20\% marginally
vignetted area. We took nine exposures of 720 s in $B_{\rm H}$ and 360
s in $R_{\rm H}$ Harris filters (a total of 6480 s and 3240 s in each
band) developing a dithering pattern of nine positions. This observing
mode allowed us to build a ``supersky'' frame that was used to correct
our images for fringing patterns (Gullixson \cite{gul92}). In
addition, the dithering helped us to clean cosmic rays and avoid gaps
between the CCDs in the final images.
Another effect associated with the wide field frames is the distortion
of the field. In order to match the photometry of several filters, a
good astrometric solution is needed to take into account these
distortions.  Using the $imcoords$ IRAF tasks and taking as a
reference the USNO B1.0 catalog, we were able to find an accurate
astrometric solution (rms $\sim$0.4\arcss) across the full frame. The
photometric calibration was performed by observing standard Landolt
fields (Landolt \cite{lan92}).

We finally identified galaxies in our $B_{\rm H}$ and $R_{\rm H}$
images and measured their magnitudes with the SExtractor package
(Bertin \& Arnouts \cite{ber96}) and AUTOMAG procedure. In a few cases
(e.g.\ close companion galaxies, galaxies close to defects of the CCD)
the standard SExtractor photometric procedure failed. In these cases
we computed magnitudes by hand. This method consists in assuming a
galaxy profile of a typical elliptical galaxy and scaling it to the
maximum observed value. The integration of this profile gives us an
estimate of the magnitude. This method is similar to PSF photometry,
but assumes a galaxy profile, more appropriate in this case.

We transformed all magnitudes into the Johnson--Cousins system
(Johnson \& Morgan \cite{joh53}; Cousins \cite{cou76}). We used
$B=B\rm_H+0.13$ and $R=R\rm_H$ as derived from the Harris filter
characterization
(http://www.ast.cam.ac.uk/$\sim$wfcsur/technical/photom/colours/) and
assuming a $B-V\sim 1.0$ for E--type galaxies (Poggianti
\cite{pog97}).  As a final step, we estimated and corrected the
galactic extinction $A_B \sim0.93$, $A_R \sim0.58$ from Burstein \&
Heiles's (\cite{bur82}) reddening maps. These values are especially
high because A2294 is immersed in a diffuse dust cloud soaring high
above the plane of our Milky Way Galaxy, and known as Polaris Dust
Nebula.

We estimated that our photometric sample is complete down to $R=22.0$
(23.0) and $B=23.5$ (24.5) for $S/N=5$ (3) within the observed field.

We assigned magnitudes to all galaxies of our spectroscopic catalog.

Table~\ref{catalogA2294} lists the velocity catalog (see also
Fig.~\ref{figottico2}): identification number of each galaxy, ID
(Col.~1); right ascension and declination, $\alpha$ and $\delta$
(J2000, Col.~2); $B$ and $R$ magnitudes (Cols.~3 and 4); heliocentric
radial velocities, ${\rm v}=cz_{\sun}$ (Col.~5) with errors, $\Delta
{\rm v}$ (Col.~6); emission lines detected in the spectra (Col.~7).

\begin{figure}
\centering 
\resizebox{\hsize}{!}{\includegraphics{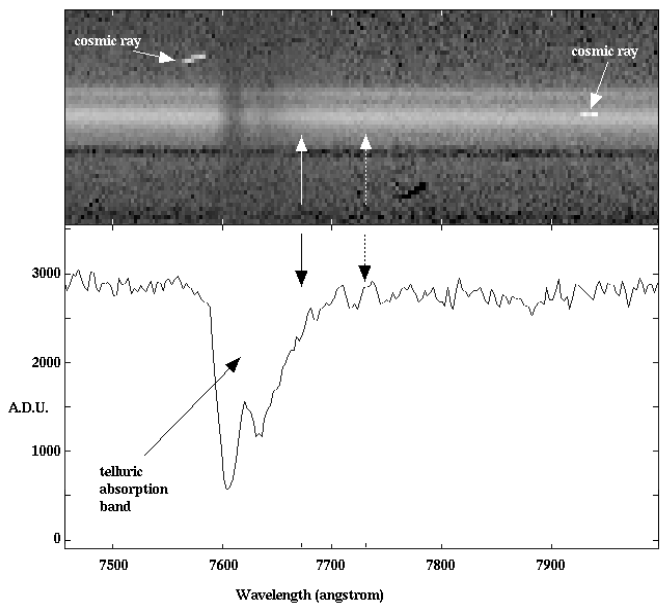}}
\caption
{{\it Top panel:} 2D spectrum of the BCG galaxy taken with the grism
  LR--R mounted on DOLORES in the wavelength range $\sim$7500--8000
  \AA. {\it Bottom panel:} 1D reduced spectrum of the BCG galaxy in the
  same wavelength range as above. Solid and dashed arrows indicate the
  position of the (hypothetical) H$\alpha$ emission line according to
  the redshift given in this paper and to the redshift provided by
  Crawford et al. (\cite{cra95}), respectively. The spectrum does not
  show any evidence of the H$\alpha$ emission.}
\label{fighalfa}
\end{figure}

The brightest galaxy of A2294 (ID.~46 in Table~\ref{catalogA2294},
hereafter BCG) is a likely dominant galaxy, 1.6 $R$--magnitudes more
luminous than other cluster members. The measured redshift is
$z=0.1690\pm0.0002$, different from that reported by Crawford et
al. (\cite{cra95}), z=0.178, using INT data and measured on the
H$\alpha$ emission line only.  This discrepancy prompted us to acquire
additional data for this galaxy. In August 2009 we acquired two 900 s
exposure long--slit spectra of the BCG. We used the LR--R grism,
covering the wavelength range $\sim$4500--10000 \AA. The target was
positioned in two slightly different positions of the slit in order to
perform an optimal sky subtraction with a technique commonly used to
reduce spectroscopic data in the near--infrared. Our reduced spectrum
(see Fig.~\ref{fighalfa}) does not show any evidence of the H$\alpha$
emission. Notice that the H$\alpha$ emission reported by Crawford
  et al. (\cite{cra95}) is very strong $\rm EW_{H\alpha}=51.7 \pm 3.2$
  (see for comparison the spectrum of A291 having $\rm
  EW_{H\alpha}=20.9 \pm 0.9$ in their Fig.~1) and thus its presence
  would be just striking in our spectrum.  Indeed, we strongly suspect
  that their detection is due to some problem in data reduction,
  e.g. a cosmic ray or sky subtraction, as also suggested by the wrong
  measure of the galaxy redshift.

Other cluster members are much less luminous than the BCG: out of
them, the brightest ones lie in the central cluster region (IDs. 28,
39 and 18) with the exception of ID. 17 (hereafter R6) which lies in
the northern region and is very radio luminous ($>10^{24}$ W
Hz$^{-1}$, No.~6 of Rizza et al. \cite{riz03}). Out of radio galaxies
listed by Rizza et al. \cite{riz03}), we have also acquired redshift
for their No.~2 (ID.~47, labelled as R2 in Fig.~\ref{figimage1}),
confirming its membership to the cluster.

\section{Analysis of the spectroscopic sample}
\label{anal}

\subsection{Member selection}
\label{memb}

\begin{figure}
\centering
\resizebox{\hsize}{!}{\includegraphics{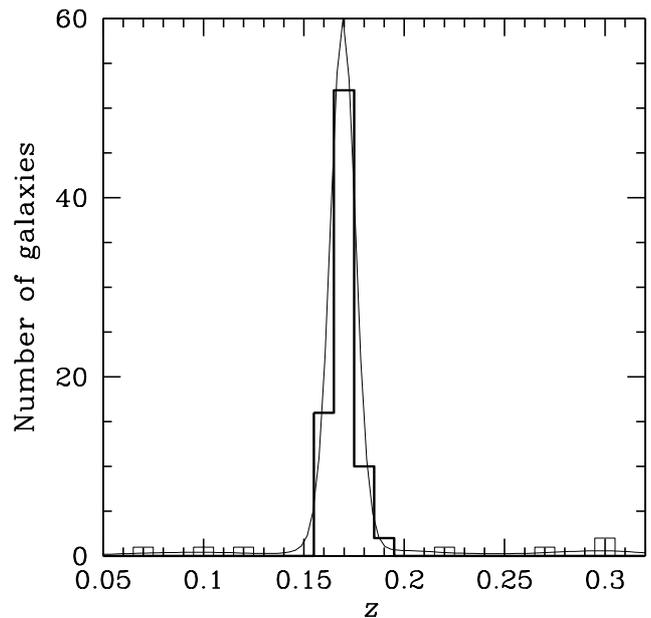}}
\caption
{Redshift galaxy distribution. The solid line histogram refers to the
  80 galaxies assigned to the A2294 complex according to the DEDICA
  reconstruction method. The number-galaxy density in the redshift
  space, as provided by the adaptive kernel reconstruction method is
  overlapped to the histogram.}
\label{fighisto}
\end{figure}

To select cluster members out of 88 galaxies having redshifts, we
follow a two steps procedure. First, we perform the 1D
adaptive--kernel method (hereafter DEDICA, Pisani \cite{pis93} and
\cite{pis96}; see also Fadda et al. \cite{fad96}; Girardi et
al. \cite{gir96}). We search for significant peaks in the velocity
distribution at $>$99\% c.l.. This procedure detects A2294 as a peak
at $z\sim0.169$ populated by 80 galaxies considered as candidate
cluster members (in the range $46553\leq {\rm v} \leq 58151$, see
Fig.~\ref{fighisto}). Out of eight non--members, three and five are
foreground and background galaxies, respectively.

\begin{figure}
\centering 
\resizebox{\hsize}{!}{\includegraphics{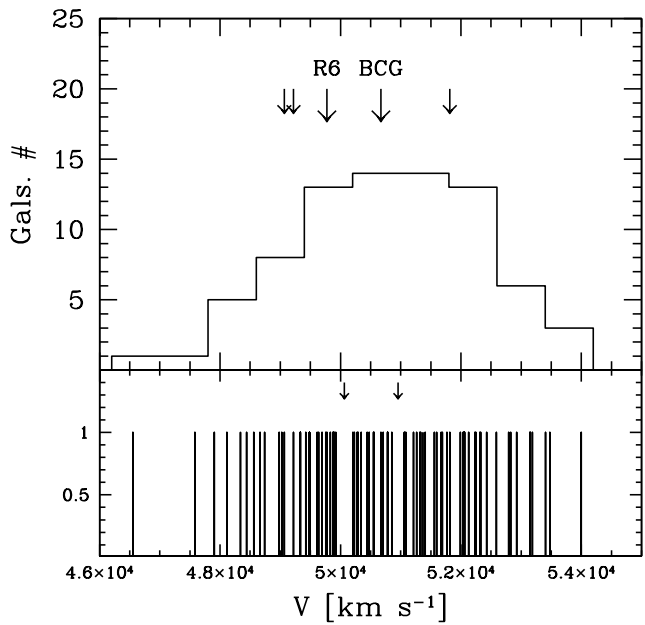}}
\caption
{The 78 galaxies assigned to the cluster.  {\em Upper panel}: Velocity
  distribution.  The arrows indicate the velocities of the five
  brightest galaxies, in particular we indicate the brightest cluster
  galaxy ``BCG'', the bright radio galaxy ``R6''.  {\em Lower panel}:
  Stripe density plot where the arrows indicate the positions of the
  significant gaps.}
\label{figstrip}
\end{figure}

All the galaxies assigned to the cluster peak are analyzed in the
second step which uses the combination of position and velocity
information: the ``shifting gapper'' method by Fadda et
al. (\cite{fad96}).  This procedure rejects galaxies that are too far
in velocity from the main body of galaxies within a fixed bin that
shifts along the distance from the cluster center.  The procedure is
iterated until the number of cluster members converges to a stable
value.  Following Fadda et al. (\cite{fad96}) we use a gap of $1000$
\ks -- in the cluster rest--frame -- and a bin of 0.6 \hh, or large
enough to include 15 galaxies. As for the center of A2294 we adopt the
position of the BCG [R.A.=$17^{\mathrm{h}}24^{\mathrm{m}}12\dotsec14$,
  Dec.=$+85\degree 53\arcmm 12\arcs$ (J2000.0)], which is almost
coincident to the X-ray centroid obtained in this paper using Chandra
data (see Sect.~\ref{xray}). The ``shifting gapper'' procedure rejects
other two obvious interlopers very far from the main body ($>2000$
\kss) but survived to the first step of our member selection
procedure.  We obtain a sample of 78 fiducial members (see
Fig.~\ref{figstrip}).

  The five member galaxies showing emission lines (ELGs) are
  preferentially found in the external cluster regions (see
  Fig.~\ref{figottico2}). The only ELG close to the cluster center
  (ID.~24) lies in the high tail of the velocity distribution, far more
  than $cz=2500$ \ks from the mean cluster velocity, as expected
  e.g. in the case of a very radial orbit.  These findings are in
  general agreement with large statistical analyses of ELGs in
  clusters (see Biviano et al. \cite{biv97} and refs. therein).

\subsection{Global cluster properties}
\label{glob}

By applying the biweight estimator to the 78 cluster members (Beers et
al. \cite{bee90}, ROSTAT software), we compute a mean cluster redshift
of $\left<z\right>=0.1693\pm$ 0.0005, i.e.
$\left<\rm{v}\right>=(50769\pm$155) \kss.  We estimate the LOS
velocity dispersion, $\sigma_{\rm V}$, by using the biweight estimator
and applying the cosmological correction and the standard correction
for velocity errors (Danese et al. \cite{dan80}).  We obtain
$\sigma_{\rm V}=1363_{-91}^{+110}$ \kss, where errors are estimated
through a bootstrap technique.

\begin{figure}
\centering
\resizebox{\hsize}{!}{\includegraphics{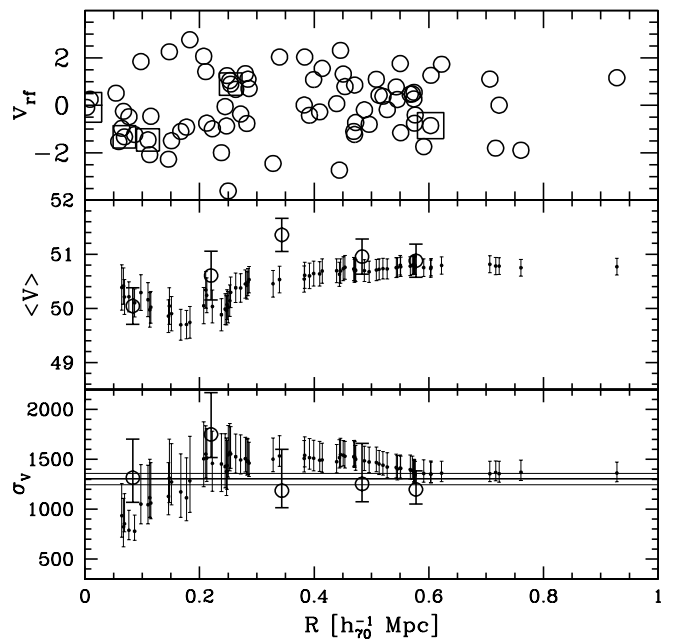}}
\caption
{{\em Top panel:} rest--frame velocity vs. projected distance from the
  cluster center.  Squares indicate the five brightest galaxies.  {\em
    Middle and bottom panels:} differential (big circles) and integral
  (small points) profiles of mean velocity and LOS velocity
  dispersion, respectively.  For the differential profiles we plot the
  values for five annuli from the center of the cluster, each of 0.25
  \h (large symbols).  For the integral profiles, the mean and
  dispersion at a given (projected) radius from the cluster--center is
  estimated by considering all galaxies within that radius -- the
  first value computed on the five galaxies closest to the center. The
  error bands at the $68\%$ c.l. are also shown.  In the lower panel,
  the horizontal line represents the X--ray temperature (see
  Sect.~\ref{xray}) with the respective errors transformed in
  $\sigma_{\rm V}$ assuming the density--energy equipartition between
  ICM and galaxies, i.e.  $\beta_{\rm spec}=1$ (see
  Sect.~\ref{disc}).}
\label{figprof}
\end{figure}

To evaluate the robustness of the $\sigma_{\rm V}$ estimate we analyze
the velocity dispersion profile (Fig.~\ref{figprof}).  The integral
profile rises out to $\sim0.2$ \h  and then flattens suggesting
that a robust value of $\sigma_{\rm V}$ is asymptotically reached in
the external cluster regions, as found for most nearby clusters (e.g.,
Fadda et al. \cite{fad96}; Girardi et al. \cite{gir96}).  

In the framework of usual assumptions (cluster sphericity, dynamical
equilibrium, coincidence of the galaxy and mass distributions), one
can compute virial global quantities. Following the prescriptions of
Girardi \& Mezzetti (\cite{gir01}), we assume for the radius of the
quasi--virialized region ${\rm R}_{\rm vir}=0.17\times \sigma_{\rm
  V}/H(z) = 3.05$ \h -- see their eq.~1 with the scaling with $H(z)$
(see also eq.~ 8 of Carlberg et al. \cite{car97} for ${\rm R}_{200}$).
We compute the virial mass (Limber \& Mathews \cite{lim60}; see also,
e.g., Girardi et al. \cite{gir98}):

\begin{equation}
M=3\pi/2 \cdot \sigma_{\rm V}^2 {\rm R}_{\rm PV}/G-{\rm SPT},
\end{equation}

\noindent where SPT is the surface pressure term correction (The \&
White \cite{the86}), and ${\rm R}_{\rm PV}$ is a projected radius (equal to
two times the projected harmonic radius).

The estimate of $\sigma_{\rm V}$ is robust when computed within a
large cluster region (see Fig.~\ref{figprof}).  The value of ${\rm
  R}_{\rm PV}$ depends on the size of the sampled region and possibly
on the quality of the spatial sampling (e.g., whether the cluster is
uniformly sampled or not).  Since in A2294 we have sampled only a
fraction of ${\rm R}_{\rm vir}$, we have to use an alternative
estimate of ${\rm R}_{\rm PV}$ on the basis of the knowledge of the
galaxy distribution. Following Girardi et al. (\cite{gir98}; see also
Girardi \& Mezzetti \cite{gir01}) we assume a King--like distribution
with parameters typical of nearby/medium--redshift clusters: a core
radius ${\rm R}_{\rm c}=1/20\times {\rm R}_{\rm vir}$ and a
slope--parameter $\beta_{\rm fit,gal}=0.8$, i.e. the volume galaxy density
at large radii goes as $r^{-3 \beta_{fit,gal}}=r^{-2.4}$. We obtain ${\rm
  R}_{\rm PV}(<{\rm R}_{\rm vir})=2.27$ \hh, where a $25\%$ error is
expected (Girardi et al. \cite{gir98}, see also the approximation
given by their eq.~13 when $A={\rm R}_{\rm vir}$). The value of SPT strongly
depends on the radial component of the velocity dispersion at the
radius of the sampled region and could be obtained by analyzing the
(differential) velocity dispersion profile, although this procedure
would require several hundred galaxies. We decide to assume a $20\%$
SPT correction as obtained in the literature by combining data on many
clusters sampled out to about ${\rm R}_{\rm vir}$ (Carlberg et
al. \cite{car97}; Girardi et al. \cite{gir98}). We compute $M(<{\rm
  R}_{\rm vir}=3.05 \hhh)=3.7_{-1.0}^{+1.1}$ \mquii.

\subsection{Velocity distribution}
\label{velo}

We analyze the velocity distribution to look for possible deviations
from Gaussianity that might provide important signatures of complex
dynamics. For the following tests the null hypothesis is that the
velocity distribution is a single Gaussian.

We estimate three shape estimators, i.e. the kurtosis, the skewness,
and the scaled tail index (see, e.g., Beers et al.~\cite{bee91}).
We find no evidence that the velocity distribution departs from Gaussianity.

Then we investigate the presence of gaps in the velocity distribution.
We follow the weighted gap analysis presented by Beers et
al. (\cite{bee91}; \cite{bee92}; ROSTAT software).  We look for
normalized gaps larger than 2.25 since in random draws of a Gaussian
distribution they arise at most in about $3\%$ of the cases,
independent of the sample size (Wainer and Schacht~\cite{wai78}). We
detect two significant gaps (at the $97\%$ and $98.6\%$ c.l.s) which
divide the cluster in three groups of 28, 14 and 36 galaxies from low
to high velocities (hereafter GV1, GV2 and GV3, see
Fig.~\ref{figstrip}).  The BCG is assigned to the GV2 peak.
Among other luminous galaxies, R6, IDs. 28 and 39 are assigned to the
GV1 peak and ID.~18 to the GV3 peak. 

Following Ashman et al. (\cite{ash94}) we also apply the Kaye's
mixture model (KMM) algorithm. This test does not find a three--groups
partition which is a significantly better descriptor of the velocity
distribution with respect to a single Gaussian.

\subsection{3D--analysis}
\label{3d}

The existence of correlations between positions and velocities of
cluster galaxies is a footprint of real substructures.  Here we use
three different approaches to analyze the structure of A2294 combining
position and velocity information.

\begin{figure}
\centering 
\resizebox{\hsize}{!}{\includegraphics{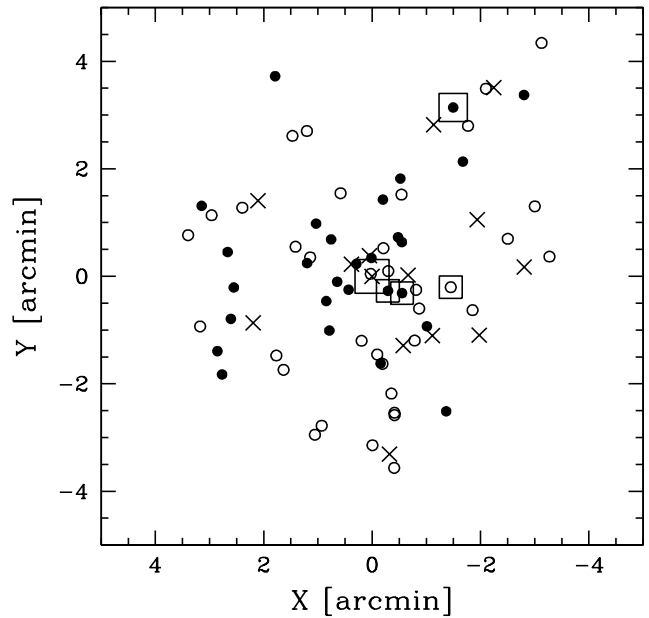}}
\caption
{Spatial distribution on the sky of the cluster galaxies showing the
  three groups recovered by the weighted gap analysis. Solid circles,
  crosses and open circles indicate the galaxies of GV1, GV2 and GV3,
  respectively.  The BCG is taken as the cluster center.  Large
  squares indicate the five brightest cluster members. Among them,
  the BCG and R6 are indicated by the two largest squares.}
\label{figxyv123}
\end{figure}

To look for a possible physical meaning of the three subclusters
determined by the two weighted gaps we compare two by two the spatial
galaxy distributions of GV1, GV2, and GV3.  We find that the GV1 and
GV3 groups  differ in the distributions of the
clustercentric distances of member galaxies at the $93\%$ c.l.
(according to the the 1D Kolmogorov--Smirnov test; hereafter
1DKS--test, see e.g., Press et al. \cite{pre92}).  The GV1 galaxies
are, on average, closer to the cluster center than the GV3 galaxies
(see Fig.~\ref{figxyv123}).

We analyze the presence of a velocity gradient performing a multiple
linear regression fit to the observed velocities with respect to the
galaxy positions in the plane of the sky (e.g, Boschin et
al. \cite{bos04} and refs. therein). We find a position angle on the
celestial sphere of $PA=214_{-28}^{+33}$ degrees (measured
counter--clock--wise from north), i.e. higher--velocity galaxies lie
in the SSW region of the cluster. To assess the significance of this
velocity gradient we perform 1000 Monte Carlo simulations by randomly
shuffling the galaxy velocities and for each simulation we determine
the coefficient of multiple determination ($RC^2$, see e.g., NAG
Fortran Workstation Handbook \cite{nag86}).  We define the
significance of the velocity gradient as the fraction of times in
which the $RC^2$ of the simulated data is smaller than the observed
$RC^2$. We find that the velocity gradient is marginally significant
at the $91\%$ c.l..

We also combine galaxy velocity and position information to compute
the $\Delta$--statistics devised by Dressler \& Schectman
(\cite{dre88}; see also e.g., Boschin et al. \cite{bos06} for a recent
application).  We find no significant indication of substructure.

\subsection{Kinematics of more luminous galaxies}
\label{segre}

The presence of velocity segregation of galaxies with respect to their
colors, luminosities, and morphologies is often taken as evidence of
advanced dynamical evolution of the parent cluster (e.g. Biviano et
al. \cite{biv92}; Fusco-Femiano \& Menci \cite{fus98}).  Here we check
for possible luminosity segregation of galaxies in the velocity space.

We find no significant correlation between the absolute velocity
$|{\rm v}|$ and $R$--magnitude.  We also divide the sample in a low
and a high--luminosity subsamples by using the median
$R$--magnitude=18.145. The two subsamples do not differ in their
velocity distribution as we verify with the standard means--test and
F--test (e.g., Press et al.~\cite{pre92}) applied to the means and
variances of velocities and with the 1DKS--test applied to the whole
velocity distributions. This is in agreement with the very small range
of action of velocity segregation in galaxy clusters, i.e. typically
only the three most luminous galaxies (Biviano et al. \cite{biv92};
see also Goto \cite{got05}).

\begin{figure}
\centering
\resizebox{\hsize}{!}{\includegraphics{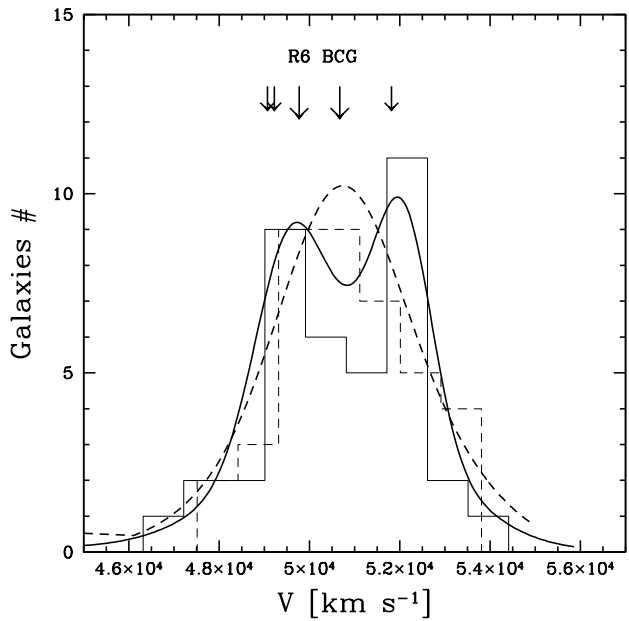}}
\caption
{Velocity galaxy distribution of more and less luminous cluster
  members (solid and dashed lines, respectively).  The arrows indicate
  the velocities of the five brightest galaxies, in particular ``BCG''
  indicates the bright, central galaxy and ``R6'' indicates the bright
  radio--galaxy.  The number-galaxy densities, as provided by the adaptive
  kernel reconstruction method are overlapped to the histograms.}
\label{figdensisegre}
\end{figure}

Examining the velocity distributions of the two subsamples in more
details we find that the distribution of luminous galaxies is found to
be non--Gaussian according to the scale tail index (at the $95\%$
c.l.)  and that, according to the 1D--DEDICA technique, it is better
described by a bimodal distribution (see
Fig.~\ref{figdensisegre}). The two peaks of this distribution, of 20
and 19 galaxies at $\sim 49700$ and 51950 \ks respectively, are
separated by $\sim 2000$ \ks in the rest cluster frame and appear
largely overlapped, i.e. 15 galaxies have a non--null probability to
belong to both the peaks. The BCG is assigned to the low--velocity
peak, but with a large probability to belong to the other peak. Among
other luminous galaxies, R6, IDs.~28 and 39 are assigned to the
low--velocity peak and ID.~18 to the high--velocity peak.

According to the DEDICA assignment, we estimate $\sigma_{\rm
    V}\sim 780$ and $\sim 640$ \ks for the low and high-velocity
  groups, respectively.  However, since there is a wide
  velocity--range where galaxies have a non--zero probability of
  belonging to both the clumps, DEDICA membership assignment leads to
  an artificial truncation of the of the tails of the distributions.  This
  truncation may lead to an underestimate of velocity dispersion for
  the subclusters. Thus, we prefer to rely on the estimates obtained
  through the KMM approach even if the best bimodal fit is not a
  significant improvement with respect to the single Gaussian
  according to the likelihood ratio test.  The low-- and
  high--velocity groups given by the best KMM bimodal fit have mean
  velocities $\left<\rm{v}\right>\sim49740$ and 52230 \kss, in good
  agreement with the peak velocities reported above and $\sigma_{\rm
    V}\sim 1070$ and $\sim 670$ \kss.

\subsection{Analysis of the photometric sample}
\label{photo}

\begin{figure}
\centering
\resizebox{\hsize}{!}{\includegraphics{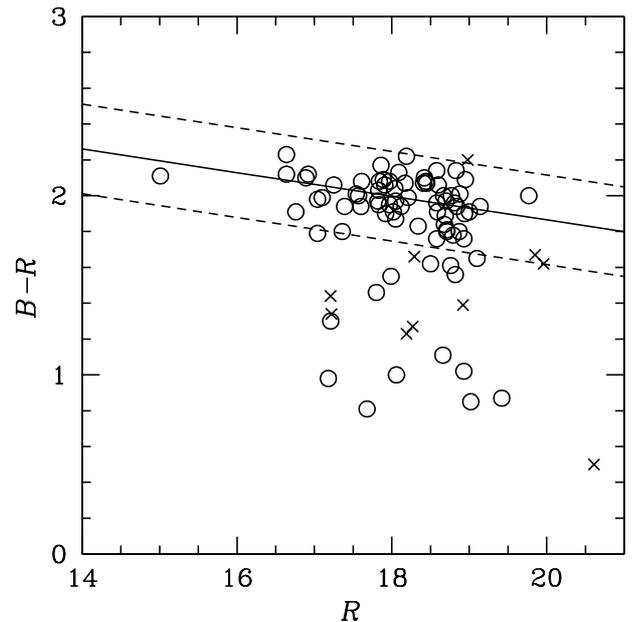}}
\caption
{$B$--$R$ vs. $R$ diagram for galaxies with available spectroscopy is shown
by circles and crosses (cluster and field members, respectively).
The solid line gives the best--fit color--magnitude relation as
determined on member galaxies; the dashed lines are drawn at $\pm$0.25
mag from the CMR. 
}
\label{figcm}
\end{figure}

\begin{figure}
\centering
\resizebox{\hsize}{!}{\includegraphics{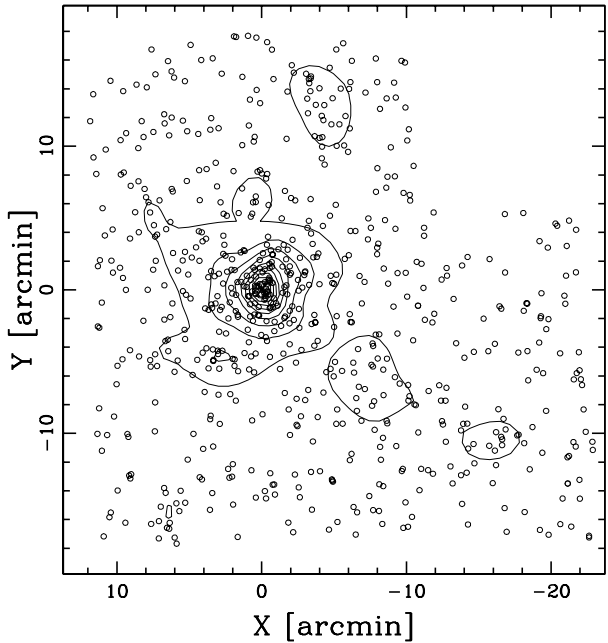}}
\caption
{Spatial distribution on the sky and relative isodensity contour map
  of likely cluster members extracted from our photometric catalog
  with $R\le 21$. The contour map is obtained with the DEDICA method
  (black lines). The plot is centered on the cluster center.}
\label{figk2}
\end{figure}

By applying the 2D adaptive--kernel method to the positions of A2294
galaxy members we find only one significant peak.  However, our
spectroscopical data do not cover the entire cluster field and suffer
of magnitude incompleteness.  To overcome these limits we recover our
photometric catalog selecting likely members on the basis of the
color--magnitude relation (hereafter CMR), which indicates the
early--type galaxy locus.  To determine the CMR we fix the slope
according to L\'opez--Cruz et al. (\cite{lop04}, see their Fig.~3) and
apply the two--sigma--clipping fitting procedure to the cluster
members obtaining $B$--$R=3.185-0.066\times R$ (see Fig.~\ref{figcm}).
Out of our photometric catalog we consider as likely cluster members
those objects with a SExtractor stellar index $\le 0.9$ lying within
0.25 mag of the CMR.  To avoid contamination by field galaxies we do
not show results for galaxies fainter than 21 mag (in $R$--band).
Figure~\ref{figk2} shows the contour map for the likely cluster
members having $R\le 21$: we find again that A2294 is well described
by only one peak centered on the BCG galaxy. 

\section{X--ray analysis}
\label{xray}

The X--ray analysis of A2294 is performed on the archival data of the
Chandra ACIS--I observation 800246 (exposure ID \#3246). The pointing
has an exposure time of 10 ks. Data reduction is performed using the
package CIAO\footnote{CIAO is freely available at
  http://asc.harvard.edu/ciao/} (Chandra Interactive Analysis of
Observations, ver. 3.3 with CALDB ver. 3.2.1) on chips I0, I1,
I2 and I3 (field of view $\sim 17\arcmin\times 17\arcmin$). First, we
remove events from the level 2 event list with a status not equal to
zero and with grades one, five and seven. Then, we select all events
with energy between 0.3 and 10 keV. In addition, we clean bad offsets
and examine the data, filtering out bad columns and removing times
when the count rate exceeds three standard deviations from the mean
count rate per 3.3 s interval. We then clean the four chips for
flickering pixels, i.e., times where a pixel has events in two
sequential 3.3 s intervals. The resulting exposure time for the
reduced data is 9.9 ks.

\begin{figure}
\centering
\resizebox{\hsize}{!}{\includegraphics{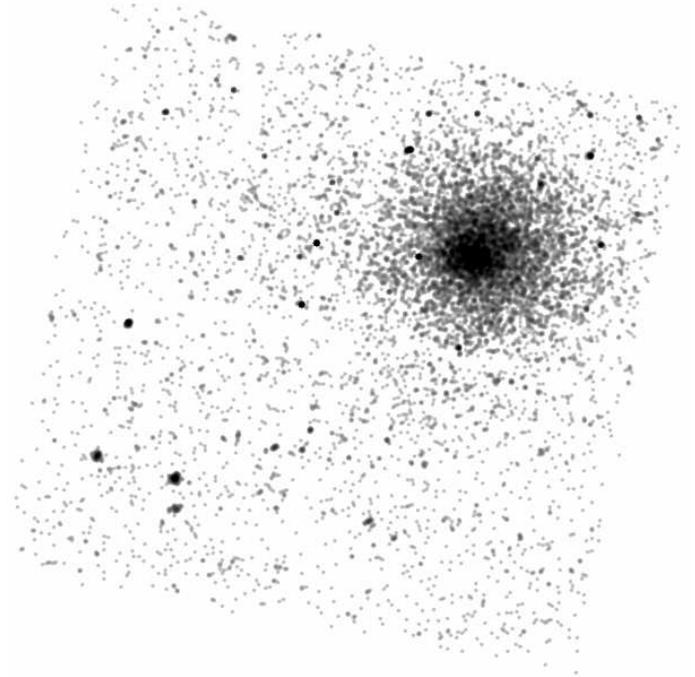}}
\caption
{17\arcmm$\times$17\arcm Chandra X--ray smoothed image (ID~3246) of
  A2294 in the energy band 0.5--2 keV (North at the top and East to
  the left).}
\label{figX}
\end{figure}

A quick look at the reduced image is sufficient to reveal the regular
morphology of the extended X--ray emission of this cluster (see
Fig.~\ref{figX}). The low values of the $P_{\rm m}/P_0$ power ratios
found by Bauer et al. (\cite{bau05}) quantitatively support this
feeling. The absence of multiple clumps in the ICM is confirmed by
performing a wavelet multiscale analysis on the chip I3. In fact, the
task CIAO/Wavdetect identifies A2294 as a single extended X--ray
source.

To better characterize the X--ray morphology of the cluster, by using
the CIAO package Sherpa we fit a simple Beta model to the 2D X--ray
photon distribution on the chip I3. The model is defined as follows (Cavaliere \&
Fusco--Femiano \cite{cav76}):

\begin{equation}
S({\rm R})=S_0[1+({\rm R}/{\rm R}_{\rm c})^2]^{\alpha}+b,
\end{equation}
where $\rm R$ is the projected radial coordinate from the centroid
position and $b$ the surface brightness background level. Before the
fit we binned the image by a factor 8 and divided it by a normalized
exposure map. The best fit centroid position is located at
R.A.=$17^{\mathrm{h}}24^{\mathrm{m}}04\dotsec87$ and Dec.=$+85\degree
53\arcmm 15.6\arcs$ (J2000.0, with an error of $\pm 0.8\arcss$) at
$\sim$8.5\arcs from the position of the BCG. The slope parameter is
$\alpha$=-0.96$\pm$0.05, that is $\beta_{\rm
  fit,gas}=(-\alpha+0.5)/3=0.49\pm 0.02$, and the (angular) core
radius is R$_{\rm c}=36.2\arcs_{-2.5\arcs}^{+2.7\arcs}$. At the
redshift of A2294 the core radius corresponds to 104.4$^{+7.7}_{-7.3}$
\kpcc.

\begin{figure}
\centering
\resizebox{\hsize}{!}{\includegraphics{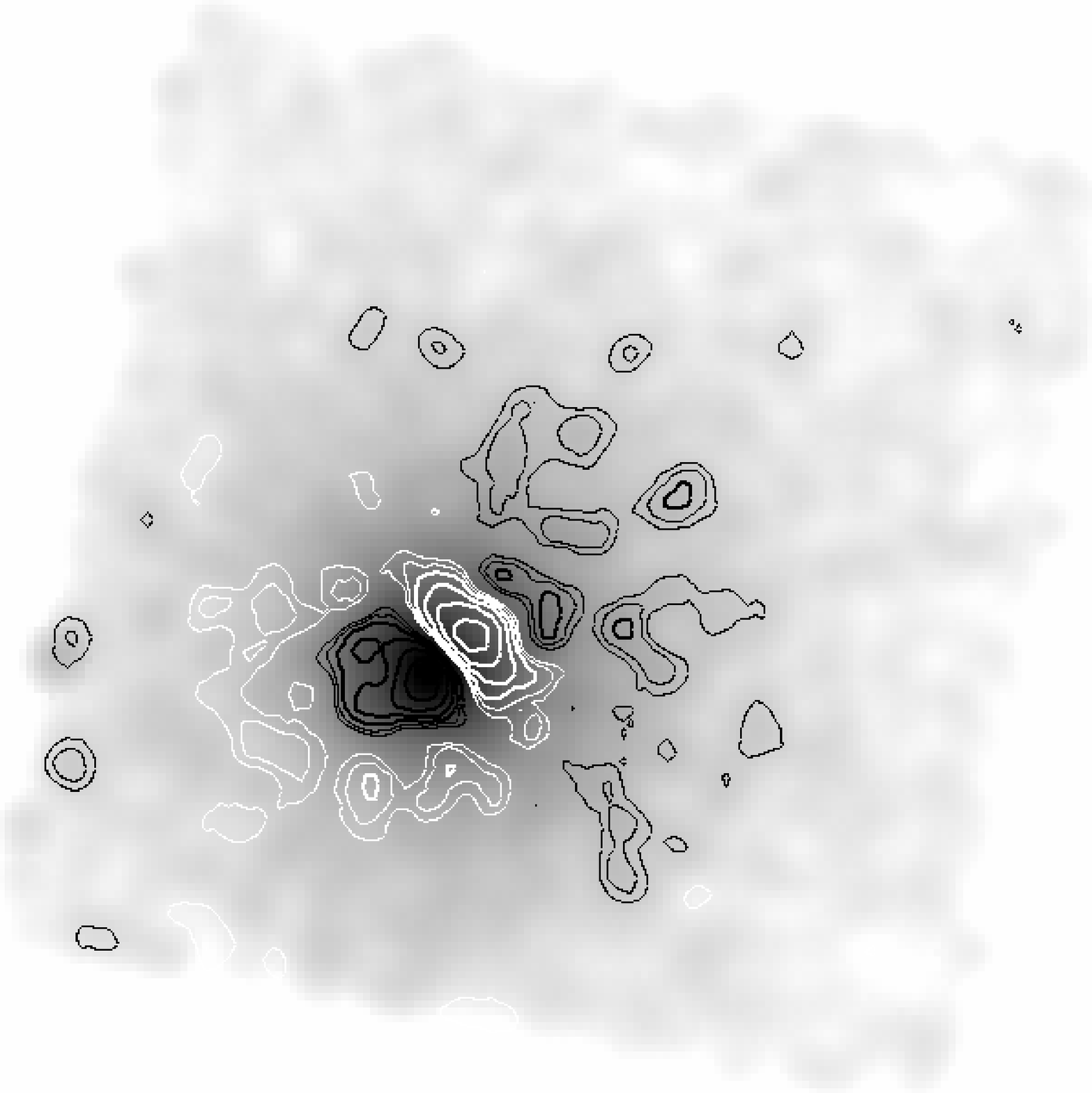}}
\caption
The smoothed X--ray emission in the right-upper quadrant of
  Fig.~\ref{figX} with, superimposed, the contour levels of the
  positive (black) and negative (white) smoothed Beta model residuals
  (North at the top and East to the left).
\label{residui}
\end{figure}

The above model is an adequate description fit to the data (the
  reduced CSTAT statistic is 1.04; Cash \cite{cas79}). However, we
check for possible departures of the X--ray surface brightness, and
thus of the gas density distribution, from the Beta model fit by
investigating the Beta model residuals. The residuals show a deficit
of X--ray emitting gas in a region extending along the NE--SW
direction, with a negative peak in the very central cluster region
(see Fig.~\ref{residui}). Around the cluster center, elongated in the
direction SE--NW, there are also regions with excess of positive
residuals.

As for the spectral properties of the cluster X--ray photons, we
compute a global estimate of the ICM temperature. The temperature is
computed from the X--ray spectrum of the cluster within a circular
aperture of $\sim$173\arcs radius (0.5 \h at the cluster redshift)
around the centroid of the X--ray emission. Fixing the absorbing
galactic hydrogen column density at 6.19$\times$10$^{20}$ cm$^{-2}$,
computed from the HI maps by Dickey \& Lockman (\cite{dic90}), we fit
a Raymond--Smith (\cite{ray77}) spectrum using the CIAO package Sherpa
with a $\chi^{2}$ statistics and assuming a metal abundance of 0.3 in
solar units. We find a best fitting temperature of $T_{\rm
  X}=\,$10.3\,$\pm\,1.1$ keV. 

A detailed temperature and metallicity map would be highly desirable
to better describe the properties of the ICM, but due to the
relatively small exposure time the photon statistics is not good
enough to this aim. However, a low SNR option to detect possible
  temperature gradients in the ICM is producing a ``hardness'' (or
  ``softness'') map of the cluster. We create two images in the energy
  bands 0.5--2 keV (soft band) and 2--7 keV (hard band), subtracting a
  constant background level in each energy band. Computing counts in
  the soft ($S$) and in the hard ($H$) band we define the quantity
  ``softness ratio'' as $SR=(S-H)/(S+H)$. Both images are
  exposure-corrected with their corresponding exposure maps. The low
  number of photons available force us to choose a large pixel size in
  order to get a good count statistic per pixel, so the resolution of
  the soft and hard images is low (182 kpc pix$^{-1}$). A 3D graph of
  the $SR$ values in the pixels within a radius of 0.5 \h from the
  cluster center is reported in Fig.~\ref{softvshard}. Typical errors
  are $\pm 0.4$, while the median value of $SR$ is 0.25. According to
  the PIMMS tool this value corresponds to a temperature of $\sim 9.7$
  keV, in good agreement with the global temperature reported
  above. The $SR$ 3D graph does not show any evidence of significant
  temperature gradients.

\onlfig{13}{
\begin{figure}
\centering
\includegraphics[width=18cm]{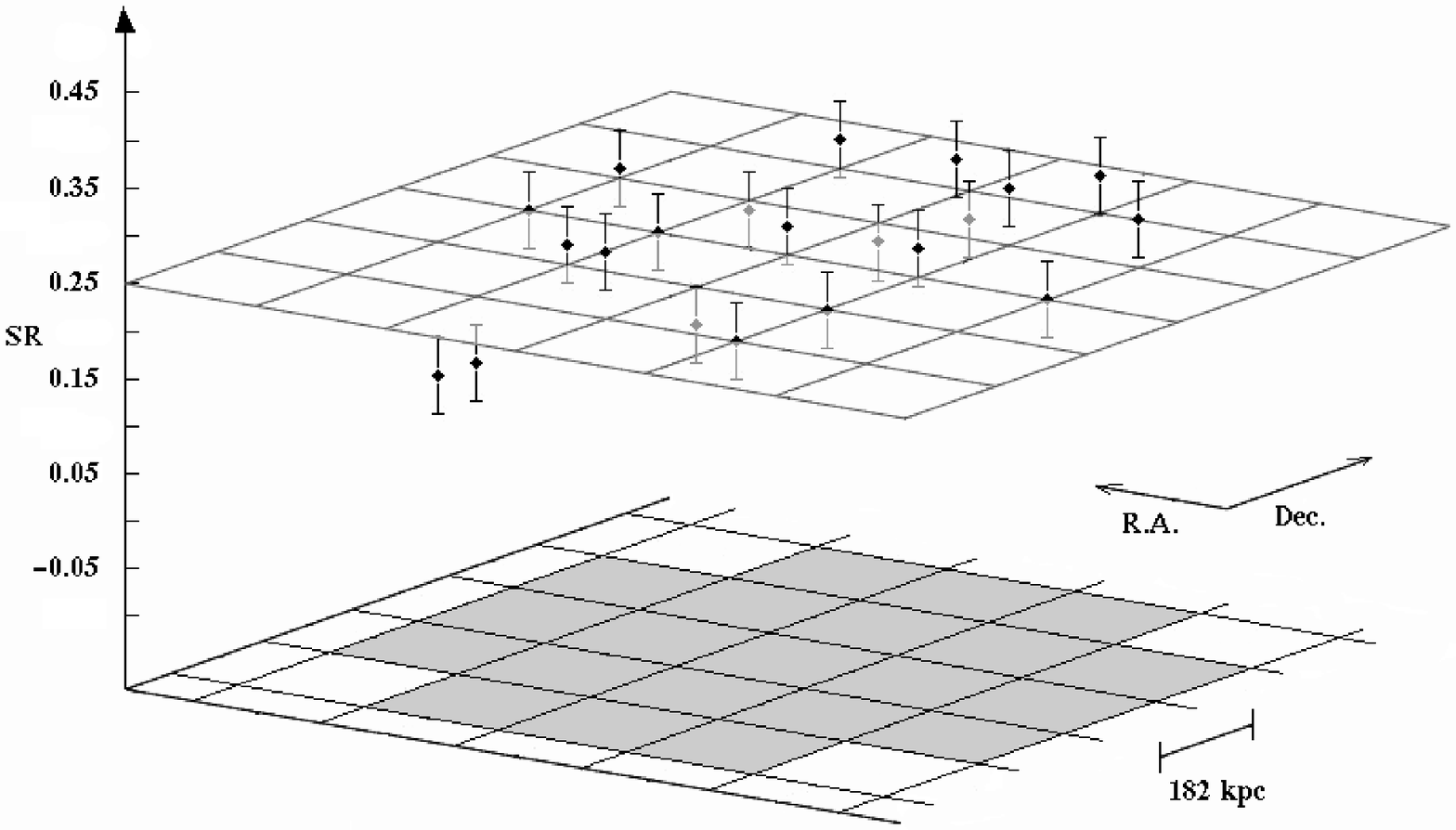}
\caption
{Softness ratio 3D map of A2294 around the centroid of the X--ray
distribution. The plane at the median $SR$ value (0.25) is drawn.}
\label{softvshard}
\end{figure}
}

\section{Discussion and conclusions}
\label{disc}

Our estimate of the cluster redshift is $\left<z\right>=0.1693\pm$
0.0005 and the BCG is well at rest within the cluster (cfr. our
$z=0.1690$ with $z=0.178$ by Crawford et al. \cite{cra95}).

For the first time the internal dynamics of A2294 is analyzed.  

The high values of the velocity dispersion $\sigma_{\rm
  V}=1363_{-91}^{+110}$ \ks and X--ray temperature $T_{\rm
  X}=(10.3\pm1.1)$ keV are comparable to the highest values found in
typical clusters (see Mushotzky \& Scharf \cite{mus97}; Girardi \&
Mezzetti \cite{gir01}; Leccardi \& Molendi \cite{lec08}).  Our
estimates of $\sigma_{\rm V}$ and $T_{\rm X}$ are fully consistent
when assuming the equipartition of energy density between ICM and
galaxies. In fact, we obtain $\beta_{\rm spec} =1.09^{+0.18}_{-0.15}$
to be compared with $\beta_{\rm spec}=1$, where $\beta_{\rm
    spec}=\sigma_{\rm V}^2/(kT/\mu m_{\rm p})$ with $\mu=0.58$ the
  mean molecular weight and $m_{\rm p}$ the proton mass (see also
Fig.~\ref{figprof}). Taking on face this result one might think that
A2294 is not far from dynamical equilibrium and consider very reliable
the virial mass estimate $M(<{\rm R}_{\rm vir}\sim 3\hhh)\sim 4$ \mqui
computed in \S~\ref{glob}.

\subsection{Internal structure}
\label{disc1}

However, our analysis shows signs that this cluster is not so relaxed
as one can think at a first glance. Evidence in this sense comes
from both optical and X--ray analyses.

First of all, both the integral and differential velocity dispersion
profiles rise in the central region out to $\sim0.2$ \h
(Fig.~\ref{figprof}, middle and bottom panels). As for the velocity
dispersion, this behavior might be a signature of a relaxed cluster
due to circular velocities and galaxy mergers phenomena in the central
cluster region (e.g., Merritt \cite{mer88}; Menci \& Fusco
Femiano~\cite{men96}; Girardi et al.~\cite{gir98}). Alternatively, it
might be due to the presence of subclumps, having different mean
velocities (see, e.g., Abell 3391--3395 in Girardi et
al. \cite{gir96}; Abell 115 in Barrena et al. \cite{bar07b}). The
latter hypothesis is supported by the behavior of the mean velocity
profile and by the plot of velocity vs. projected clustercentric
distance where the central region is more populated by low velocity
galaxies than high velocity galaxies.  This suggests the presence of
substructure in the cluster core.

Second, the two gaps found in the velocity distribution suggest the
presence of three subclumps with the BCG in the middle velocity
subclump. Although the presence of the three groups (GV1, GV2, GV3) is
not very strongly significant on the base of only velocity data, the
existence of a spatial segregation between GV1 and GV3 groups is a
footprint of real substructures. We also find a (marginally
significant) velocity gradient toward the SSW direction.

Third, the high--luminosity galaxy subsample shows two peaks (largely
overlapped) in the velocity distribution, with the BCG someway in the
middle. This result is very appealing, since galaxies of different
luminosity could trace the dynamics of cluster mergers in a different
way. A noticeable example was reported by Biviano et
al. (\cite{biv96}): they found that the two central dominant galaxies
of the Coma cluster are surrounded by luminous galaxies, accompanied
by the two main X--ray peaks, while the distribution of faint galaxies
tend to form a structure not centered with one of the two dominant
galaxies, but rather coincident with a secondary peak detected in
X--ray. Biviano et al. speculate that the merging is in an advanced
phase, where faint galaxies trace the forming structure of the
cluster, while the most luminous galaxies still trace the remnant of
the core--halo structure of a pre--merging clump, which could be so
dense to survive for a long time after the merging (as suggested by
numerical simulations, e.g.  Gonz\'alez--Casado et
al. \cite{gon94}). In A2294, we can speculate that luminous galaxies
trace the remnants of two merging subclusters characterized by an
impact velocity $\sim$ 2000 \kss.  Assuming the dynamical equilibrium
for each of the two individual subclusters, from the values of $\sigma_{\rm
  V}$ of the two subclusters we obtain a virial mass of 1.8 \mqui and
0.5 \mqui for the low and high velocity subclusters. The total
mass $M_{\rm sys}=2.3$ \mqui is lower than the global virial
value computed in \S~\ref{glob}, but still a high value.

As for X--ray data, we find no evidence of obvious substructure.  In
fact, our multiscale wavelet analysis of the Chandra image does not
reveal any subclumps in the X--ray photon distribution.  
  Moreover, we confirm the absence of a significant, macroscopic
  cluster ellipticity (see also Hashimoto et al. \cite{has07}; Maughan
  et al. \cite{mau08}).

As for the Beta model we fit, the value of the core radius
  R$_{\rm c}=104.4^{+7.7}_{-7.3}$ \kpc and the value of the slope
  parameter $\beta_{\rm fit,gas}=0.49\pm 0.03$ well agree with those
  computed by Hart in his PhD thesis (\cite{har08}; R$_{\rm
    c}=(99\pm12)$ \kpcc; $\beta_{\rm fit,gas}=0.48\pm 0.02$).  The
value of the core radius agrees with that expected from the relation
between surface brightness concentration index and core radius shown
by Hashimoto et al. (\cite{has07}, see their Fig.~8 and the value of
concentration in their Table~2).  The values of R$_{\rm c}$ and
$\beta_{\rm fit,gas}$ well lie on the low end of the parabolic
relation found between these two parameters (Neumann \& Arnaud
\cite{neu99}).  On the other hand, the value of $\beta_{\rm
  fit,gas}=0.49\pm 0.03$ might seem somewhat small considering the
typical values for very rich/hot clusters (e.g., Jones \& Forman \cite
{jon99}; Vikhlinin et al. \cite{vik99}).  However, notice that most
our signal comes from the region with a radius of $\sim 0.5$ \h ($\sim
1/6\,{\rm R_{vir}}$) and that there are indications for continuous
steepening of the X--ray brightness profiles with increasing radius
(e.g., Vikhlinin et al. \cite{vik99}; Neumann \cite{neu05}). This
steepening is the likely cause of offsets between different cluster
samples (see Vikhlinin et al. \cite{vik99} where $\beta_{\rm
  fit,gas}\gtrsim 0.6$ vs.  Jones \& Forman \cite {jon99} where
$\left<\beta_{\rm fit,gas}\right>=0.6$) and of apparent discrepancies
between fit parameters obtained for the same clusters (e.g. Buote et
al. \cite{buo05}). Indeed, the most appropriate way to compare
different clusters it seems to consider the measure of the local
  slope of the surface brightness at a certain, rescaled radius 
  (see Croston et al. \cite{cro08} for variation of this parameter
  with radius). As for A2294, Maughan et al. (\cite{mau08}) computed
the slope $\beta_{500}=1.22^{+0.32}_{-0.25}$ at a radius of
R$_{500}=1.3$ \hh, using the data in the radial range
  0.7R$_{500}$-1.3R$_{500}$, i.e. well out of the region we analyze. This
value is in agreement with that expected for very hot clusters at
$z<0.5$ (see their Fig.~11).  Finally, we notice that, in the case of
a cluster merger, numerical simulations predict a clear expansion of
the gas core and a somewhat steepening of the slope (Roettiger et
al. \cite{roe96}, see their Fig.~3).  This is in agreement with that
suggested by Jones \& Forman (\cite {jon99}) to explain the large core
radii found for a few observed clusters, but see Neumann \& Arnaud
(\cite{neu99}) for no link between $\beta_{\rm fit,gas}$ value and
cluster dynamical status. Summarizing, our small values of R$_{\rm c}$
and $\beta_{\rm fit,gas}$ are not suggestive of substructure.

Direct evidence of cluster substructure comes from the 2D image of the
Beta model residuals, which shows positive residuals in the X--ray
emission along the SE--NW direction (see Fig.~\ref{residui}).

\begin{figure}
\centering
\resizebox{\hsize}{!}{\includegraphics{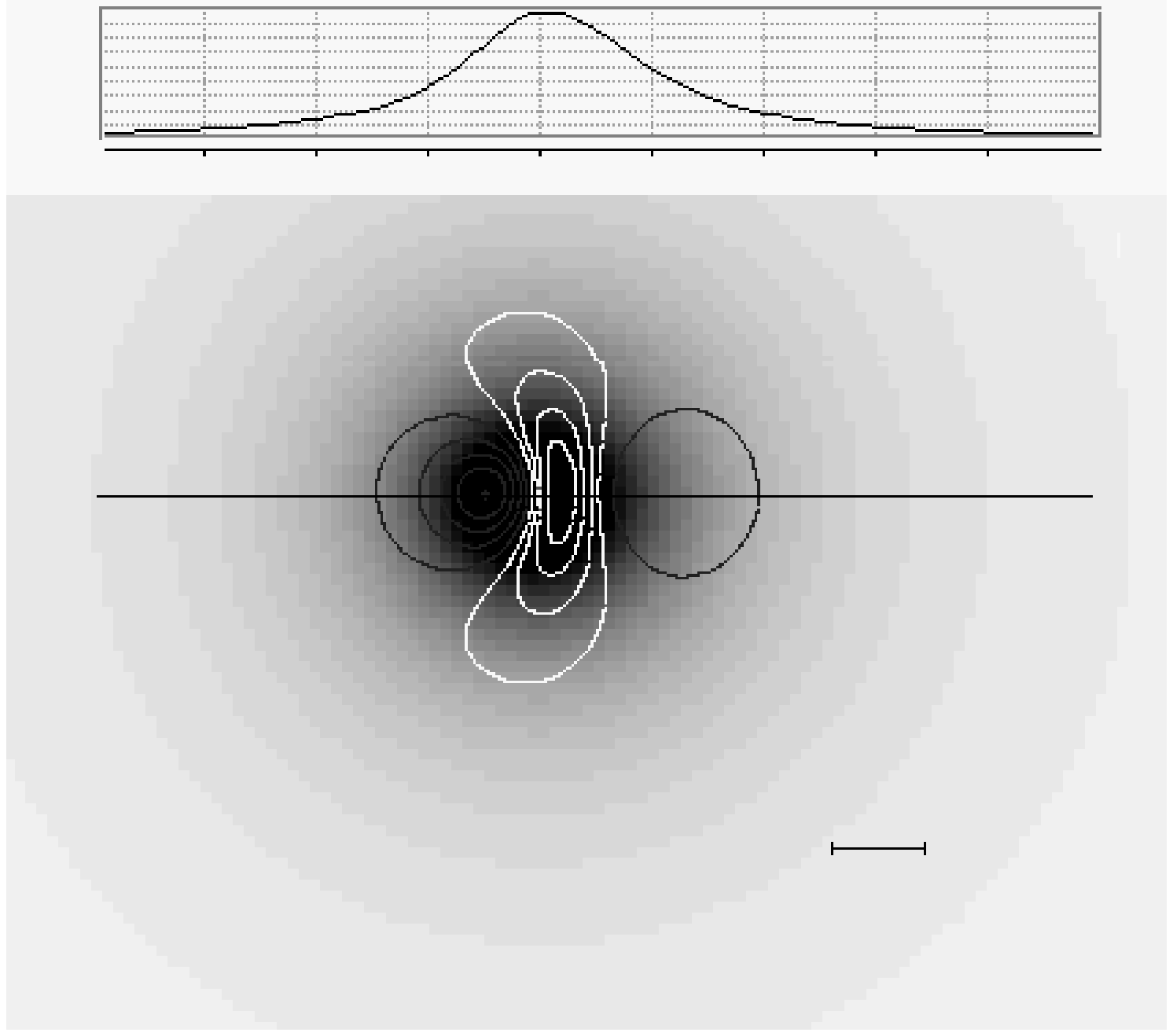}}
\caption
{{\it Upper panel:} Integrated X--ray profile of the simulated
  system composed by two subsystems, both following a Beta model
  profile but with different model parameters (see text). {\it Lower
    panel:} Surface brightness distribution of the simulated system
  with, superimposed, the contour levels of the Beta model
  residuals. White (dark/gray) contours represent negative (positive)
  residuals. As a reference, the size of the fitted core radius is
  drawn.}
\label{residuimodel}
\end{figure}

In order to interpret the residual image we simulate two systems, both
having an X--ray surface brightness profile following a Beta model
with the same $\beta_{\rm fit,gas}$, but different R$_{\rm c}$
  and $S_0$, with the centers separated by a distance of the order of
  the two adopted core radii. The surface brightness profile of the
composed system has a single peak, as in the case of A2294 (see
  Fig.~\ref{residuimodel}, upper panel). The fit with a single Beta
  model provides a value for R$_{\rm c}\sim 50\%$ and $\sim20\%$
  larger than the two adopted core radii, respectively. Instead,
  $\beta_{\rm fit,gas}$ is $\sim 10\%$ larger than the adopted value.
The appearance of the 2D image of the residuals
(Fig.~\ref{residuimodel}, lower panel) is roughly similar to that
obtained for A2294, with a two--clump surplus of X--ray photons 
  (with the left clump being the most evident) in the line defined by
the centers of the subsystems and a deficit in the perpendicular
direction (cfr. Fig.~\ref{residuimodel} with Fig.~\ref{residui}). Thus
the residual image of A2294 data might be explained by the presence of
two very close (or very closely projected) systems along the SE--NW
direction.  In particular, we find that an asymmetry between the
  two components might explain the more prominent excess of the SE
  structure in the residual image. Some pieces of observational
evidence found in the literature, i.e.  the presence of a centroid
shift (Rizza et al. \cite{riz98} and Maughan et al. \cite{mau08}, but
see Hashimoto et al. \cite{has07}) or a certain degree of asymmetry of
the X-ray profile around the centroid of the photon distribution
(Hashimoto et al. \cite{has07}) are likely amenable to the
substructure we detect.  Obviously, our above bimodal model is a very
simplified scenario in the case of a close interaction between two
galaxy systems -- see the section below --and we do not attempt to go
further in interpreting observed data, e.g exploring in more detail
their quantitative parameters.

$\ $

\subsection{Investigating the likely merger cluster}
\label{disc2}

The absence of a macroscopic elongation of the galaxy and ICM
distributions and the poor significance of the velocity
gradient suggests that the evidence of substructure we detect is
a trace of minor/old accretion phenomena or that the direction of the
cluster merger is aligned with the LOS. A LOS merger direction
would make more difficult the analysis of the cluster internal dynamics.
Another example of a cluster merger along the LOS is the galaxy
cluster CL~0024+17, an apparently relaxed system, which is actually a
collision of two clusters, the interaction occurring along our LOS, as
shown by about 300 redshifts in the cluster field (Czoske et
al. \cite{czo02} and refs. therein).  

The cluster merger scenario is generally consistent with the
  absence of the cool core. In fact, although simulations yield
  ambivalent results about the role of mergers in destroying cool
  cores (Poole et al. \cite{poo06}; Burns et al. \cite{bur08}),
  observations seem to favor cool core destruction through cluster
  mergers (Allen et al. \cite{all01}; Sanderson et
  al. \cite{san06}). In particular, the LOS merging direction might
  explain the high compactness of A2294 with respect to other
non--cool core clusters (Bauer et al. \cite{bau05}, see their Fig.~3).
Also notice that our new data on the BCG exclude the presence of
H$\alpha$ emission previously reported by Crawford et
al. (\cite{cra95}), thus reconducting A2294 to a quite ``normal''
non--cool core.

In the framework of a cluster merger where the two subclusters
  are well traced by the luminous galaxies (for the non--collisional
  part, i.e. dark matter and galaxies) and the residual image (for the
  collisional part, i.e. the gas), we can also obtain some information
  about the evolutionary stage of the merger.  Assuming that
  $\beta_{\rm spec}=1$ for each of the two subclusters, from the
  values of $\sigma_{\rm V}$ we obtain the X--ray temperatures $T_{\rm
    X}=$7.0 and 2.8 keV.  The observed X--ray temperature is thus
  $\sim 1.4$ times that of the main subcluster. While the observed
  X--ray temperature of the merging simulated clusters is still not
  clear at later times (e.g. 2-3 Gyr after the collision, see ZuHone
  et al. \cite{zuh09} and refs. therein), numerical simulations agree
  in finding enhancements of the X--ray temperature around the time of
  the cross core. After a very sharp rise the temperature peaks just
  during the core-crossing or just after and then declines (Ricker \&
  Sarazin \cite{ric01}; Mastropietro \& Burkert \cite{mas08}).  Since
  we do not see evidence for a very hot, arc-shaped feature in the
  cluster center, we assume that the merger is catched after the
  core-crossing, i.e.  in the outgoing phase.  For the case of a 1:3
  mass ratio, Fig.~8 of Ricker \& Sarazin (\cite{ric01}) suggests a
  time $\lesssim 0.5$ Gyr after the core-crossing.

At this point we have the minimum observation-based information to
apply the two-body model (Beers et al. \cite{bee82}; Thompson
\cite{tho82}) following the methodology outlined for, e.g., Abell 1240
(Barrena et a. \cite{bar09}).  This simple model assumes radial orbits
for the clumps with no shear or net rotation of the system. According
to the boundary conditions usually considered, the clumps are assumed
to begin their evolution at time $t_0=0$ with a separation $d_0=0$,
and are now moving apart or coming together for the first time in
their history.  In the case of a collision, we assume that the time
$t_0=0$ with separation $d_0=0$ is the time of their core crossing and
that we are looking at the system a time $t$ after.  The values of the
relevant parameters for the two--subclusters system are $t\sim 0.5$
Gyr; the relative LOS velocity in the rest--frame, $V_{\rm rf}\sim
2000$ \kss; the projected linear distance between the two clumps,
$D\sim 0.1$ \hh.  The last parameter is deduced from the residual
image and thus might be an underestimate for the non-collisional
component.

\begin{figure}
\centering
\resizebox{\hsize}{!}{\includegraphics{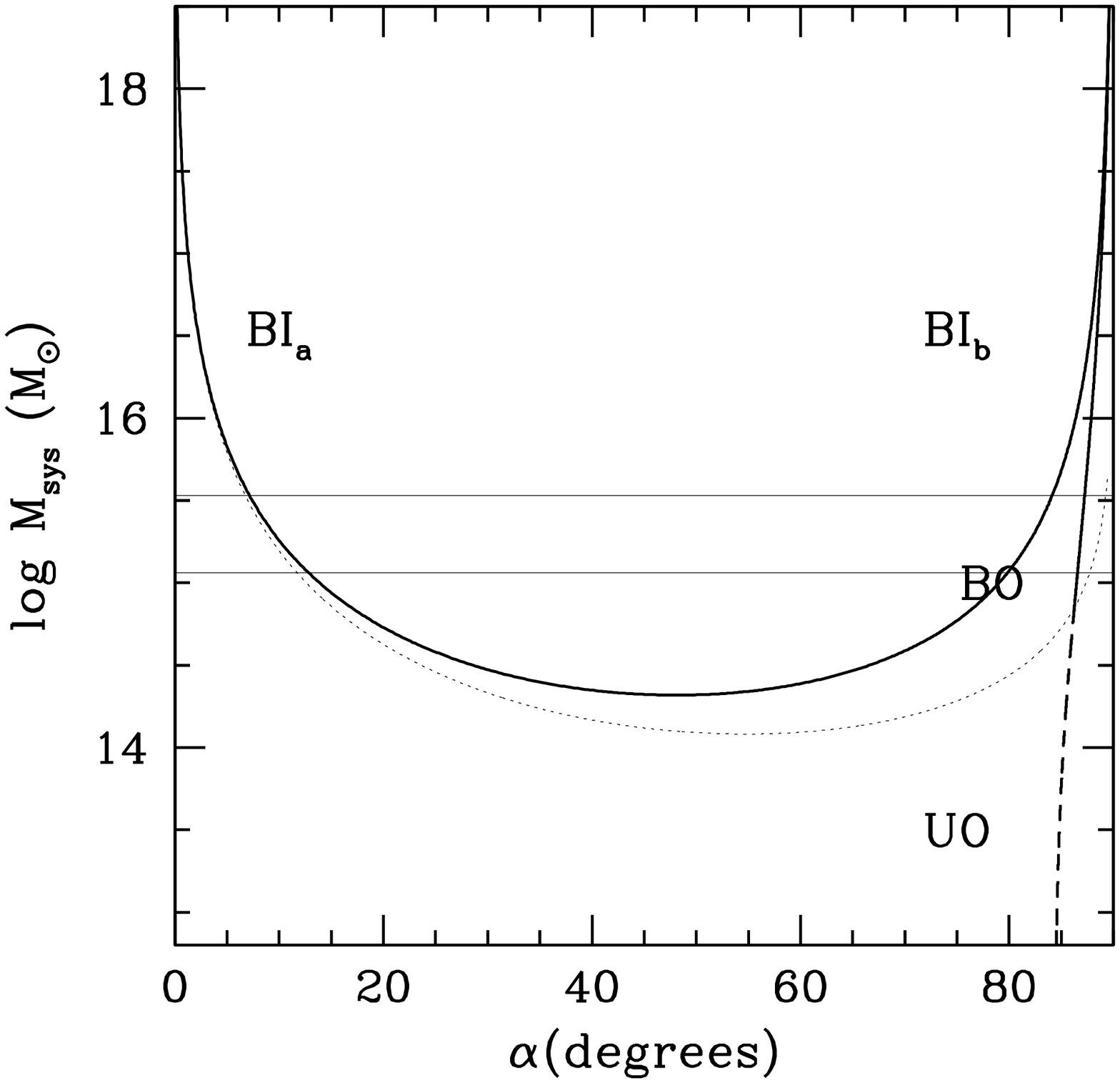}}
\caption
{System mass vs. projection angle for bound and unbound solutions
  (thick solid and thick dashed curves, respectively) of the two--body
  model applied to the low and high velocity subclusters. Labels
  BI$_{\rm a}$ and BI$_{\rm b}$ indicate the bound and incoming, i.e.,
  collapsing solutions (thick solid curve). Label BO indicates the
  bound outgoing, i.e., expanding solutions (thick solid curve). Label
  UO indicates the unbound outgoing solutions (thick dashed
  curve). The horizontal lines give the range of observational values
  of the mass system with a 50\% error. The thin dashed curve
  separates bound and unbound regions according to the Newtonian
  criterion (above and below the thin dashed curve, respectively).}
\label{figbim}
\end{figure}

The bimodal model solution gives the total system mass $M_{\rm sys}$,
i.e. the sum of the masses of the two subclusters, as a function of
$\alpha$, where $\alpha$ is the projection angle between the plane of
the sky and the line connecting the centers of the two clumps (e.g.,
Gregory \& Thompson \cite{gre84}).  Figure~\ref{figbim} compares the
bimodal--model solutions with the observed mass of the system $M_{\rm
  sys}=2.3$ \mqui considering a 50\% uncertainty band.  Among other
solutions, we find the bound outgoing solution (BO) with $\alpha \sim
85$, i.e.  the cluster merger is occurring largely in the LOS
direction, in agreement with our expectations.  In the framework
  of this solution the SE clump, which is the more X--ray luminous
  and thus likely the more massive,
  is moving towards SE in the direction of the observer, while the
  less X--ray luminous and massive NW subcluster is moving
  towards NW in the opposite direction with respect to the observer.
The true spatial distance between the two subclumps is $D_{\rm
  3D}\sim$ 1 \h and the real, i.e. deprojected, velocity difference is
$V_{\rm rf,3D} \sim 2000 $ \kss.  In this scenario we expect the
presence of a gas shock (Bykov et al. \cite{byk08} and refs. therein).
We can estimate the Mach number of the shock from ${\cal M}={\rm
  v}_{\rm s}/c_{\rm s}$, where ${\rm v}_{\rm s}$ is the velocity of
the shock and $c_{\rm s}$ in the sound speed in the pre--shock gas
(see e.g., Sarazin \cite{sar02} for a review). The value of $c_{\rm
  s}$ can be obtained from our estimate of $\sigma_{\rm V}\sim 1070$
for the most massive subcluster.  For the value of ${\rm v}_{\rm s}$
we use ${\rm v}_{\rm s}\gtrsim V_{\rm rf,3D}=2000 $ \kss, since after
the core crossing the shock velocity is larger than the subcluster
velocity (see Fig.~4 of Springel \& Farrar \cite{spr07} and Fig.~14 of
Mastropietro \& Burkert \cite{mas08}). This leads to ${\cal
  M}$$\gtrsim 2$, in agreement with moderate Mach numbers $2\le$${\cal
  M}$$\le4$ expected for shocks due to the cluster merging.

$\ $

In conclusion, present observational evidence is consistent with A2294
being a very massive cluster just formed or in the phase of forming
through a merger, i.e.  similar to most DARC clusters we previously
analyzed (e.g. Boschin et al. \cite{bos06}; Barrena et
al. \cite{bar07a}; Girardi et al. \cite{gir08}).  The time--scale
  of a few fractions of Gyr agrees both with the results of other merging
  clusters showing radio halos/relics (e.g., Markevitch et
  al. \cite{mar02}; Girardi et al.  \cite{gir08}; Barrena et
  al. \cite{bar09}) and with theoretical expectations for radio halos
  (Brunetti et al. \cite{bru09}). Instead the morphology of the A2294
  radio halo is somewhat intriguing. In fact, after the subtraction
  of discrete sources, the radio halo appears someway elongated along
  the EW direction as shown by Giovannini et al. (\cite{gio09}; their
  Fig.~10 on the left). Previously analyzed clusters of DARC sample
  showed radio halos no elongated or elongated in the direction of the
  merger (Abell 697, Girardi et al. \cite{gir06}; Abell 520, Girardi et
  al. \cite{gir08}). This appearent misaligneament with the project
  merging direction (SE--NW) deserves to be
  furtherly investigated.
  
  More in general, to verify our hypothesis about a cluster merger in
  A2294 and better quantify the merging framework we suggest both the
  acquisition of many more redshifts in the cluster field and/or
  deeper X--ray observations. In particular, X--ray data would allow
  to check the temporal phase of the merger although the LOS geometry
  of the merger would make difficult the direct observation of the
  shock (e.g., Markevitch et al. \cite{mar05}).  The acquisition of more
  redshifts might allow to better determine the non-collisional
  components of the merging subclusters.

\begin{acknowledgements}

We are in debt with Gabriele Giovannini for the VLA radio image and
his useful comments.  We thank the anonymous referee for his/her
very stimulating suggestions. This publication is based on observations
made on the island of La Palma with the Italian Telescopio Nazionale
Galileo (TNG) and the Isaac Newton Telescope (INT). The TNG is
operated by the Fundaci\'on Galileo Galilei -- INAF (Istituto
Nazionale di Astrofisica). The INT is operated by the Isaac Newton
Group. Both telescopes are located in the Spanish Observatorio of the
Roque de Los Muchachos of the Instituto de Astrofisica de Canarias.

This research has made use of the NASA/IPAC Extragalactic Database
(NED), which is operated by the Jet Propulsion Laboratory, California
Institute of Technology, under contract with the National Aeronautics
and Space Administration.

\end{acknowledgements}


\begin{thebibliography}{}

\bibitem[1989]{abe89} Abell, G. O., Corwin, H. G. Jr., \& Olowin, R. P. 1989, \apjs, 70, 1

\bibitem[2001]{all01} Allen, S. W., Ettori, S., \& Fabian, A. C. 2001, \mnras, 324, 877


\bibitem[1994]{ash94} Ashman, K. M., Bird, C. M., \& Zepf, S. E. 1994, \aj, 108, 2348


\bibitem[1994]{bar94} Bardelli, S., Zucca, E., Vettolani, G., et al. 1994, \mnras, 267, 665 

\bibitem[2007a]{bar07a} Barrena, R., Boschin, W., Girardi, M., \& Spolaor, M. 2007a, \aap, 467, 37

\bibitem[2007b]{bar07b} Barrena, R., Boschin, W., Girardi, M., \& Spolaor, M. 2007b, \aap, 469, 861

\bibitem[2009]{bar09} Barrena, R., Girardi, M., Boschin, 
\& Das\'i, M. 2009, \aap, 503, 357

\bibitem[2005]{bau05} Bauer, F. E., Fabian, A. C., Sanders, J. S., Allen, S. W., \& Johnstone, R. M. 2005, \mnras, 359, 1481
 
\bibitem[1990]{bee90} Beers, T. C., Flynn, K., \& Gebhardt, K. 1990, \aj, 100, 32

\bibitem[1991]{bee91} Beers, T. C., Forman, W., Huchra, J. P., Jones, C., \& Gebhardt, K. 1991, \aj, 102, 1581

\bibitem[1992]{bee92} Beers, T. C., Gebhardt, K., Huchra, J. P., et al. 1992, \apj, 400, 410

\bibitem[1982]{bee82} Beers, T. C., Geller, M. J., \& Huchra, J. P. 1982, \apj, 257, 23


\bibitem[1996]{ber96} Bertin, E., \& Arnouts, S. 1996, \aaps, 117, 393

\bibitem[1996]{biv96} Biviano, A., Durret, F., Gerbal, D. et al. 1996, \aap, 311, 95

\bibitem[1992]{biv92} Biviano, A., Girardi, M., Giuricin, G., Mardirossian, F., \& Mezzetti, M. 1992, \apj, 396, 35

\bibitem[1997]{biv97} Biviano, A., Katgert, P., Mazure, A. et al. 1997,
\aap, 321, 84

\bibitem[2008]{bos08} Boschin, W., Barrena, R., Girardi, M., \& Spolaor, M. 2008, \aap, 487, 33

\bibitem[2004]{bos04} Boschin, W., Girardi, M., Barrena, R., et al. 2004, \aap, 416, 839

\bibitem[2006]{bos06} Boschin, W., Girardi, M., Spolaor, M., \& Barrena, R. 2006, \aap, 449, 461

\bibitem[2009]{bru09} Brunetti, G., Cassano, R., Dolag, K., \& Setti, G. 2009, \aap, 507, 661

\bibitem[2002]{buo02} Buote, D. A. 2002, in ``Merging Processes in
Galaxy Clusters'', eds. L. Feretti, I. M. Gioia, \& G. Giovannini (The
Netherlands, Kluwer Ac. Pub.): Optical Analysis of Cluster Mergers

\bibitem[2005]{buo05} Buote, D. A., Humphrey, P. J., \& Stocke, T.
2005, \apj, 630, 750

\bibitem[2008]{bur08} Burns, J. O., Hallman, E. J., Gantner, B., Motl, P. M., \& Norman, M. L. 2008, \apj, 675, 1125

\bibitem[1982]{bur82} Burstein, D., \& Heiles, C. 1982, \aj, 87, 1165

\bibitem[2008]{byk08} Bykov, A. M., Dolag, K., \& Durret, F.	
2008, Space Science Reviews, Volume 134, Issue 1-4, pp. 119-140 

\bibitem[1997]{car97} Carlberg, R. G., Yee, H. K. C., \& Ellingson, E. 1997, \apj, 478, 462

\bibitem[1979]{cas79} Cash, W. 1979, \apj, 228, 939

\bibitem[2005]{cas05} Cassano, R., \& Brunetti, G. 2005, \mnras, 357, 1313

\bibitem[2009]{cas09} Cassano, R., \& Brunetti, G., R\"ottgering, H. J. A., \& Br\"uggen, M. 2009, \aap, in press (preprint arXiv:0910.2025v1)

\bibitem[1976]{cav76} Cavaliere, A. \& Fusco--Femiano, R. 1976, \aap, 49, 137 

\bibitem[1976]{cou76} Cousins, A. W. J., 1976, Mem. R. Astr. Soc, 81, 25

\bibitem[1995]{cra95} Crawford, C. S., Edge, A. C., Fabian, A. C., et al. 1995, \mnras, 274, 75

\bibitem[2008]{cro08} Croston, J. H., Pratt, G. W., B\"ohringer, 
et al. 2008, \aap, 431, 443

\bibitem[2002]{czo02} Czoske, O., Moore, B., Kneib, J.--P., \& Soucail, G. 2002, \aap, 386, 31

\bibitem[1980]{dan80} Danese, L., De Zotti, C., \& di Tullio, G. 1980, \aap, 82, 322

\bibitem[1990]{dic90} Dickey, J. M., \& Lockman, F. J. 1990, \araa, 28, 215

\bibitem[1988]{dre88} Dressler, A., \& Shectman, S. A. 1988, \aj, 95, 985

\bibitem[1998]{ebe98} Ebeling, H., Edge, A. C., B\"ohringer, H., et al. 1998, \mnras, 301, 881

\bibitem[2007]{edw07} Edwards, L. O. V., Hudson, M. J., Balogh, M. L., \& Smith, R. J. 2007, \mnras, 379, 100

\bibitem[1994]{ell94} Ellingson, E., \& Yee, H. K. C. 1994, \apjs, 92, 33

\bibitem[1998]{ens98} Ensslin, T. A., Biermann, P. L., Klein, U., \& Kohle, S. 1998, \aap, 332, 395

\bibitem[2001]{ens01} Ensslin, T. A., \& Gopal--Krishna 2001, \aap, 366, 26

\bibitem[1996]{fad96} Fadda, D., Girardi, M., Giuricin, G., Mardirossian, F., \& Mezzetti, M. 1996, \apj, 473, 670

\bibitem[1999]{fer99} Feretti, L. 1999, MPE Report No. 271

\bibitem[2002a]{fer02a} Feretti, L. 2002a, The Universe at Low Radio
  Frequencies, Proceedings of IAU Symposium 199, held 30 Nov -- 4 Dec
  1999, Pune, India. Edited by A. Pramesh Rao, G. Swarup, and
  Gopal--Krishna, 2002., p.133

\bibitem[2005]{fer05} Feretti, L. 2005, X--Ray and Radio Connections
(eds. L. O. Sjouwerman and K. K. Dyer). Published electronically by
NRAO, http://www.aoc.nrao.edu/events/xraydio. Held 3--6 February 2004
in Santa Fe, New Mexico, USA

\bibitem[2008]{fer06} Feretti, L. 2006, Proceedings of the XLIst
  Rencontres de Moriond, XXVIth Astrophysics Moriond Meeting: "From
  dark halos to light", L.Tresse, S. Maurogordato and J. Tran Thanh
  Van, Eds, e--print astro--ph/0612185

\bibitem[2006]{fer08} Feretti, L. 2008, Mem. SAIt, 79, 176

\bibitem[2002b]{fer02b} Feretti, L., Gioia I. M., and Giovannini
G. eds., 2002b, Astrophysics and Space Science Library, vol. 272, 
``Merging Processes in Galaxy Clusters'', Kluwer Academic Publisher,
The Netherlands

\bibitem[2008]{ferr08} Ferrari, C., Govoni, F., Schindler, S., Bykov, A. M., \& Rephaeli, Y. 2008, \ssr, 134, 93

\bibitem[1998]{fus98} Fusco--Femiano, R., \& Menci, N. 1998, \apj, 498, 95

\bibitem[2009]{gio09} Giovannini, G., Bonafede, A., Feretti, L., et al.  2009, \aap, 507, 1257

\bibitem[2002]{gio02} Giovannini, G., \& Feretti, L. 2002, in
``Merging Processes in Galaxy Clusters'', eds. L. Feretti,
I. M. Gioia, \& G. Giovannini (The Netherlands, Kluwer Ac. Pub.):
Diffuse Radio Sources and Cluster Mergers

\bibitem[1999]{gio99} Giovannini, G., Tordi, M., \& Feretti, L. 1999, New Astronomy, 4, 141

\bibitem[2007]{gir07} Girardi, M., Barrena, R., \& Boschin, W. 2007,
  Contribution to ``Tracing Cosmic Evolution with Clusters of
  Galaxies: Six Years Later'' conference --
  http://www.si.inaf.it/sesto2007/contributions/Girardi.pdf

\bibitem[2008]{gir08} Girardi, M., Barrena, R., Boschin, W., \& Ellingson, E. 2008, \aap, 491, 379

\bibitem[2002]{gir02} Girardi, M., \& Biviano, A. 2002, in ``Merging
Processes in Galaxy Clusters'', eds. L. Feretti, I. M. Gioia, \&
G. Giovannini (The Netherlands, Kluwer Ac. Pub.): Optical Analysis of
Cluster Mergers

\bibitem[2006]{gir06} Girardi, M., Boschin, W., \& Barrena, R. 2006, \aap, 455, 45

\bibitem[1996]{gir96} Girardi, M., Fadda, D., Giuricin, G. et al. 1996, \apj, 457, 61

\bibitem[1998]{gir98} Girardi, M., Giuricin, G., Mardirossian, F., Mezzetti, M., \& Boschin, W. 1998, \apj, 505, 74

\bibitem[2001]{gir01} Girardi, M., \& Mezzetti, M. 2001, \apj, 548, 79

\bibitem[2005]{got05} Goto, T. 2005, \mnras, 359, 141

\bibitem[1994]{gon94} Gonz\'alez--Casado, G., Mamon, G. A., \& Salvador--Sol\'e, E. 1994, \apjl, 433, 61

\bibitem[1984]{gre84} Gregory, S. A., \& Thompson, L. A. 1984, \apj, 286, 422

\bibitem[1992]{gul92} Gullixson, C. A. 1992, in ``Astronomical CCD Observing and Reduction techniques'' (ed. S. B. Howell), ASP Conf. Ser., 23, 130

\bibitem[2008]{har08} Hart, B. 2008, PhDT, arXiv e--print 0801.4093

\bibitem[2007]{has07} Hashimoto, Y., B\"ohringer, H., Henry, J. P., Hasinger, G., \& Szokoly, G. 2007, \aap, 467, 485

\bibitem[2004]{hoe04} Hoeft, M., Br\"uggen, M., \& Yepes, G. 2004, \mnras, 347, 389

\bibitem[1953]{joh53} Johnson, H. L., \&  Morgan, W. W. 1953, \apj, 117, 313

\bibitem[1999]{jon99} Jones, C., \& Forman, W. 1999, \apj, 511, 65

\bibitem[2004]{kem04} Kempner, J. C., Blanton, E. L., Clarke, T. E.
  et al. 2004, Proceedings of the conference ``The Riddle of Cooling
  Flows in Galaxies and Clusters of Galaxies'', eds. T. H. Reiprich,
  J. C. Kempner, \& N. Soker, e--print arXiv astro--ph/0310263

\bibitem[1992]{ken92} Kennicutt, R. C. 1992, \apjs, 79, 225

\bibitem[1992]{lan92} Landolt, A. U. 1992, \aj, 104, 340

\bibitem[2008]{lec08} Leccardi, A., \& Molendi, S. 2008, \aap, 486, 359

\bibitem[1960]{lim60} Limber, D. N., \& Mathews, W. G. 1960, \apj, 132, 286

\bibitem[2004]{lop04} L\'opez--Cruz, O., Barkhouse, W. A., \& Yee, H. K. C. 2004, \apj, 614, 679

\bibitem[1992]{mal92} Malumuth, E. M., Kriss, G. A., Dixon, W. Van Dyke, Ferguson, H. C., \& Ritchie, C. 1992, \aj, 104, 495

\bibitem[2002]{mar02} Markevitch, M., Gonzalez, A. H., David, L., et al. 2002, \apjl, 567, 27

\bibitem[2005]{mar05} Markevitch, M., Govoni, F., Brunetti, G, \& Jerius, D. 2005, \apj, 627, 733

\bibitem[2008]{mas08} Mastropietro, C., \& Burkert, A. 2008, \mnras, 389, 967

\bibitem[2008]{mau08} Maughan, B. J., Jones, C., Forman, W., \& Van Speybroeck, L. 2008, \apjs, 174, 117

\bibitem[1988]{mer88} Merritt, D. 1988, in ``The Minneosota lectures on
  clusters of galaxies and large--scale structure'' (A90--36758
  15--90). San Francisco, CA, Astronomical Society of the Pacific,
  1988, p. 175--196.

\bibitem[1996]{men96} Menci, N., \& Fusco--Femiano, R. 1996, \apj, 472, 46

\bibitem[1997]{mus97} Mushotzky, R. F., \& Scharf, C. A. 1997, \apj, 482, L13

\bibitem[1986]{nag86} NAG Fortran Workstation Handbook, 1986 (Downers Grove, IL: Numerical Algorithms Group)

\bibitem[1999]{neu99} Neumann, D. M., \& Arnaud,  M. 1999, \aap, 348, 711

\bibitem[2005]{neu05} Neumann, D. M., \& Arnaud,  M. 2005, \aap, 439, 465

\bibitem[2004]{ota04} Ota, N., Pointecouteau, E.; Hattori, M.; \& Mitsuda, K. 2004, \apj, 601, 1200	

\bibitem[1999]{owe99} Owen, F., Morrison, G., \& Voges, W. 1999, proceedings of the workshop ``Diffuse Thermal and Relativistic Plasma in Galaxy Clusters'', eds. H. B\"ohringer, L. Feretti, \& P. Schuecker, MPE Report 271, pp. 9--11

\bibitem[1998]{per98} Peres, C. B, Fabian, A. C., Edge, A. C., et al. 1998, \mnras, 298, 416

\bibitem[1993]{pis93} Pisani, A. 1993, \mnras, 265, 706

\bibitem[1996]{pis96} Pisani, A. 1996, \mnras, 278, 697

\bibitem[1997]{pog97} Poggianti, B. M. 1997, \aaps, 122, 399

\bibitem[2006]{poo06} Poole, G. B., Fardal, M. A., Babul, A., et al. 2006, \mnras, 373, 881

\bibitem[1992]{pre92} Press, W. H., Teukolsky, S. A., Vetterling, W. T., \& Flannery, B. P. 1992, in Numerical Recipes (Second Edition), (Cambridge University Press)

\bibitem[2000]{qui00} Quintana, H., Carrasco, E. R., \& Reisenegger, A. 2000, \aj, 120, 511

\bibitem[1977]{ray77} Raymond, J. C., \& Smith, B. W. 1977, \apjs, 35, 419

\bibitem[2001]{ric01} Ricker, P. M., \& Sarazin, C. L. 2001,
\apj, 561, 621

\bibitem[1998]{riz98} Rizza, E., Burns, J. O., Ledlow, M. J. et al. 1998, \mnras, 301, 328

\bibitem[2003]{riz03} Rizza, E., Morrison, G. E., Owen, F. N., et al. 2003, \aj, 126, 119

\bibitem[1996]{roe96} Roettiger, K., Burns, J. O., \& 
Loken, C. 1996, \apj, 473, 651

\bibitem[1999]{roe99} Roettiger, K., Burns, J. O., \& Stone, J. M. 1999, \apj, 518, 603

\bibitem[1997]{roe97} Roettiger, K., Loken, C., \& Burns, J. O. 1997, \apjs, 109, 307

\bibitem[2006]{san06} Sanderson, A. J. R., Ponman, T. J., \& O'Sullivan, E. 2006, \mnras, 372, 1496

\bibitem[2002]{sar02} Sarazin, C. L. 2002, in ``Merging Processes in
Galaxy Clusters'', eds. L. Feretti, I. M. Gioia, \& G. Giovannini (The
Netherlands, Kluwer Ac. Pub.): The Physics of Cluster Mergers

\bibitem[2007]{spr07} Springel, V., \& Farrar, G. R. 2007, \mnras, 380, 911

\bibitem[1986]{the86} The, L. S., \& White, S. D. M. 1986, \aj, 92, 1248

\bibitem[1982]{tho82} Thompson, L. A. 1982, in IAU Symposium 104,
  Early Evolution of the Universe and the Present Structure,
  eds. G.O. Abell and G. Chincarini (Dordrecht: Reidel)

\bibitem[1979]{ton79} Tonry, J., \& Davis, M. 1979, \apj, 84, 1511

\bibitem[1993]{tri93} Tribble, P. C. 1993, \mnras, 261, 57

\bibitem[1999]{vik99} Vikhlinin, A., Forman, W., \& Jones, C.
1999, \apj, 525, 47

\bibitem[1978]{wai78} Wainer, H., \&  Schacht, S. 1978, Psychometrika, 43, 203

\bibitem[2009]{zuh09} ZuHone, J. A., Ricker, P. M., Lamb, D. Q., \& Karen Yang, H.--Y. 2009, \apj, 699, 1004

\end{thebibliography}
\end{document}